\documentclass[11pt]{article}
\usepackage{graphics}
\usepackage{epsfig}

\newcommand{\be}{\begin{eqnarray}}
\newcommand{\ee}{\end{eqnarray}}

\def\ll#1{\left#1}
\def\r#1{\right#1}
\def\fr{\frac{1}{2}}

\def\mref#1{(\ref{#1})}

\def\p{\partial}

\def\bd{\begin{displaymath}}
\def\ed{\end{displaymath}}

\def\ba#1{\begin{array}{#1}}
\def\ea{\end{array}}
\def\nn{\nonumber}
\newfont{\Bbb}{msbm10 scaled 1200}

\begin{document}

\pagestyle{empty}

\begin{center}

{\LARGE\bf Wave Functions and Energy Spectra in Rational Billiards Are Determined Completely by Their Periods\\[0.5cm]}

\vskip 12pt

{\large {\bf Stefan Giller}}

\vskip 3pt

Jan D{\l}ugosz University in Czestochowa\\
Department of Experimental and Applied Physics\\
Armii Krajowej 13/15, 42-200 Czestochowa, Poland\\
e-mail: stefan.giller@ujd.edu.pl
\end{center}

\vspace {10pt}

\begin{abstract}
The rational billiards (RB) are classically pseudointegrable (P.J. Richens, M.V. Berry, Physica D2, 495 (1981), Stefan Giller, arXiv: 1912.04155 [quant-ph]),
i.e. their trajectories in the phase space lie on multi-tori of a genus $g$
defined by $2g$ independent periods. Each such a multi-torus can be unfolded into elementary polygon pattern (EPP) - a smallest system of mirror images of RB
obtained by their consecutive reflections by their sides and containing all different images of RB. A rational billiards Riemann surface (RBRS)
corresponding to each RB is then an infinite mosaic made by a periodic distribution of EPP. Periods of RBRS are directly related to periodic orbits
of RB. It is shown that any stationary solutions (SS) to the Schr\"odinger
equation (SE) in RB can be extended
on the whole RBRS. The extended stationary wave functions (ESS) are then periodic on RBRS with its periods. Conversely, for each system of boundary
conditions (i.e. the Dirichlet or the the Neumann ones or their mixture) consistent with EPP one can find so called stationary pre-solutions (SPS) of the Schr\"odinger
equation defined on RBRS and respecting its periodic structure together with their energy spectra. Using SPS one can easily construct SS of RB for most boundary conditions on it by a
trivial algebra over SPS. It proves therefore that the energy spectra defined by the boundary conditions for SS corresponding to each RB are totally
determined by $2g$ independent periods of RBRS being homogeneous functions of these periods. RBRS can be considered as a classical construction but in fact it
can be done exclusively due to the rationality of the polygon billiards considered. Therefore the approach developed in the present paper can be seen as a new way in obtaining
SS to SE in RB. On the other hand our results can be considered also as a generalization on the pseudointegrable systems of the well known semiclassical
result corresponding to the integrable rational billiards being however general and exact. SPS can be constructed explicitly
for a class of RB which EPP can be decomposed into a set of periodic orbit channel (POC) parallel to each other (POCDRB). For such a class of RB in which
POCs find their natural place in the quantization of RB the respective
RBRS can be built as a standard multi-sheeted Riemann surface (finitely sheeted in the case of doubly rational billiards (DRPB) and infinitely sheeted in other ones) with
a periodic structure. For POCDRB a discussion of the existence of the superscar states (SSS) (Heller, E.J., {\it Phys. Rev. Lett.} {\bf 53}, 1515 (1984);
Bogomolny E. and Schmit C., {\it Phys. Rev. Lett.} {\bf 92} 244102 (2004)) can be done thoroughly.
\end{abstract}

\vskip 10pt
\begin{tabular}{l}
{\small PACS number(s): 03.65.-w, 03.65.Sq, 02.30.Jr, 02.30.Lt, 02.30.Mv} \\[1mm]
{\small superscar states}\\[1mm]
\end{tabular}

\newpage

\pagestyle{plain}

\setcounter{page}{1}

\section{Introduction}

\hskip+2em Polygon billiards (PB) play some distinguished role in looking for relations between the quantum and the classical motions due to a
relative simplicity of these motions in the billiards. The classical motions in PB which are simply free motions of a ball between its subsequent
elastic collisions with the billiards boundaries represent a wide
variety of them starting from a small number of PB with the integrable motions and including an uncountable number of them where
the motions are chaotic. However between these two extremes there is still in the set of all PB a countable but dense set of them called rational
(RB) with all their angles being
rational (in the $\pi$-units) in which the classical motions are nor integrable nor chaotic but living in the phase space on multitori. Such classical
systems have been called pseudointegrable \cite{8}. They have the properties which makes the investigations of relations between the classical and
the quantum motions easier although do not exhaust the problem excluding chaotic PB.

Just because of their relative simplicity RB attracted particular attention of many mathematicians who were able to establish many important results on
the classical motion in RB (see e.g. Gutkin \cite{11}, Rozikov \cite{12} for a wide review of the respective results and cited papers).

Quantum properties of RB have
been investigated mainly to get an insight into the statistical properties of their energy spectra \cite{8} as well as to understand some
properties of their stationary wave functions observed as the superscar states \cite{3,6}.

In several of our earlier papers \cite{1}-\cite{5} we have studied the quantization of the pseudointegrable systems formed by RB by
the semiclassical approach to the problem. In this approach the construction of elementary polygon pattern (EPP) on which the semiclassical wave
functions have
been built has been used widely. This approach has been also applied both to RB \cite{1}-\cite{3} and \cite{5} as well as to some chaotic ones - the
Bunimovich stadium \cite{4} and the Sinai-like billiards \cite{5} approximating the latter billiards by RB ones. The results obtained by the method
have been however strongly limited by its nature since it has used plane waves
as the main elements which the semiclassical wave functions were superposed of. Therefore the method was able to describe approximately only some
parts of the energy spectra of the considered billiards and only in its high energy regions. On the other hand RBRS, a second global semiclassical construction
which appeared in the papers mentioned, did not play any important role in getting the results of the papers whereas just this notion seems to be no less
fundamental than EPP which RBRS is composed of for constructing semiclassical wave functions. However, what is much more important and will be shown in
the present paper, RBRS can allow us for
constructing a method of obtaining both the wave functions and the energy spectra in RB not appealing to the semiclassical approximations mentioned.

Let us therefore remind shortly the main properties of EPP and RBRS the constructions of which have been described in great details in \cite{5}.

First let us note that in our paper by RB is meant each multi connected billiards which outer boundaries
are formed by sides of a rational polygon inside which are holes of polygon forms too which are also rationals and all these polygons are "rotated"
with respect to the outer "mother" polygon by rational angles. EPP for such RB is done from all its possible but different images got by mirror
reflections of RB on its sides. Glueing each two images along the respective side reflecting them we get some form of EPP. Boundaries of EPP are then formed
by pairs of parallel sides belonging to two images of EPP being the mirror reflections of each other on the sides. Identifying the sides in each such
a pair of them one gets
EPP as a closed two dimensional surface topologically equivalent to two dimensional closed Riemann surface with $g$
holes in it i.e. a multi-torus of a genus $g$ in the 4-dimensional phase space of the billiards. Any two boundary points of EPP identified in the
above way are joined by a period on RBRS so that there are at least as
many periods as pairs of such parallel sides. Among these periods there are $2g$ linear independent in the space of integers. These periods are
directly related to the periodic orbits of RB considered and lengths of the orbits coincide with the ones of the periods or are integer multiples of them.

It is important to notice the local structure of EPP. If one fixes some of the vertex of RB then a part of EPP or sometimes EPP itself is developed
around this vertex by
successive mirror reflections of RB by its two sides having the vertex as their common point. If the sides form an angle equal to $p/q$
(in $\pi$-units) then there are $2q$ possible different reflections of RB around the point after which any of its sides is rotated by
the angle $2p$. If $p>1$ then such a vertex becomes a branching point for EPP on the plane, i.e. all reflected RB lie on $p$ planes with the vertex
as their common point.

RBRS appears now by gluing repeatedly the identical EPP along their boundaries made by sides of RB joined by respective periods so that two EPP
glued along some side of the boundaries are
the mirror images of each other by the side gluing them, i.e. EPP is invariant on the mirror reflection operations on any of its sides. One can
consider also this operation as repeatedly shifting one of the glued EPP by its periods.

Continuing infinitely this procedure one gets a surface with infinitely many branching points periodically distributed on it.
Rare cases of RB for which $p=1$ for all their angles are classically integrable and for
them RBRS is simply a plane. None of RB with holes belongs to this class.

It is now important to note that RBRS formed in the above way has periodic structure with the periods defined by EPP and any classical trajectory in the
billiards is transformed into RBRS as a straight line. Each branching point of RBRS being an image of a vertex with an angle
$p/q,\;p>1$, is just a bifurcation point for these trajectories which splits each
straight line trajectory passing it onto $p$ of them running by $p$ different sheets but preserving a direction of the split trajectory.
Trajectories passing by the branching points are called singular diagonals.

It should be stressed farther that the branching properties of the RB vertices on RBRS precisely describe their role in the classical and quantum descriptions of
motions in RB which otherwise is described as the strong diffraction on them of both the trajectories and the wave functions respectively. In the
classical case each vertex with $p>1$ splits a bundle of parallel trajectories into several of them running on the respective sheets generated by
the vertex while in the quantum description a wave function in RB are formed by respective superpositions of the
ones defined on all the sheets of RBRS associated with the vertex.

Let us note further that each segment of any periodic orbit of RB between its two subsequent reflections from the billiards boundary has a direction of some of the periods of
RBRS while a length of the orbit coincides with the length of the respective period or is its integer multiple.
Since any periodic orbit in RB is not
isolated it is accompanied by a continuum of other periodic trajectories parallel to it and having the same lengths and this continuum is limited by the
closest vertices of RB with $p>1$. Conversely any continuum of non-singular trajectories in RB bounded by two singular diagonals and parallel to some period of EPP
must be build of periodic trajectories too. The periodic orbit continuum is known as a periodic orbit channel (POC) \cite{6}.
Periodic trajectories of any such a continuum nowhere bifurcate running through RBRS. All other trajectories in the rational billiards
are isolated as a singular diagonal separating two neighbor POCs or form sets of parallel singular trajectories and of aperiodic ones which are densely mixed with
each other. The latter continua of parallel trajectories will be called  aperiodic orbit channel (AOC) in the paper.

It should be however stressed that building EPP and RBRS there is no need to appeal to the classical property of motions in the polygon billiards -
a possibility of their constructions follows exclusively as a result of the rationality of the considered polygon billiards, i.e. of their geometry.

It is an open question whether each RBRS can be always decomposed into a set of planes (sheets) equipped with cuts by which the planes can be glued
together making RBRS connected in the standard way known for example from the complex analysis. However if it is possible then the respective sheets
must be always glued of parallel POCs and the possibility of making such a gluing
is the necessary condition for the possible decompositions of RBRS into sheets. At the same time the examples of RB considered in the paper show also
that an existence of POCs accompanied by AOCs parallel to them does not prevent the respective RBRS to be decomposed into sheets. Of course in
such a case sheets of the respective RBRS are glued from other POCs with periods not parallel to any AOC. It follows therefore that POCs will always
play a principal role in our construction of both the periodic RBRS and the respective solutions to the stationary SE.

It is therefore justified for our further discussion to distinguish from all RB a class of them for which their RBRS can be built of sheets each of
which is glued of parallel POCs stressing also in this way the natural role played by these classical objects in the RB quantization noticed first by Bogomolny and
Schmit \cite{6}. Such RB in the paper will be called POC decomposed (POCDRB).

The properties of POCDRB discussed above and the examples of them considered in the paper justify also the following further decomposition of
POCDRB into another two classes
\begin{enumerate}
\item the class of the doubly rational polygon billiards (DRPB) \cite{2} that contains RB for which any three periods of their corresponding EPP are
linear related on the plane by rational coefficients; and
\item the class which contains all the remaining POCDRB.
\end{enumerate}

Then the following two properties seem to be valid for any DRPB
\begin{itemize}
\item its corresponding RBRS is built of a finite number of sheets; and
\item AOCs parallel to some POC are absent on such RBRS.
\end{itemize}

For the second class of RB the respective properties seem to be following
\begin{itemize}
\item their RBRS are built of infinitely many sheets; and
\item POCs on RBRS can be accompanied by AOCs.
\end{itemize}

In our earlier papers SS in RB were built on EPP in the semiclassical approximations. In some cases the latter provided us with the exact SS but mostly
we got only approximate solutions to the stationary Schr\"odinger equation and to the respective energy spectra in the high energy
regions. In this paper we would like to show that the energy eigenvalue problems for RB can be substituted by respective solutions of the Schr\"odinger
equation on RBRS rather than on EPP by looking for such solutions which respect topology of RBRS together with its periodic structure. To make such a
substitution real we will show that the exact solutions of the eigenvalue problem of the Schr\"odinger equation in RB can be continuously extended on the
whole RBRS respecting its periodic structure proving in this way the existence on RBRS of such periodic solutions to the stationary Schr\"odinger
equation. Trying to invert the extension procedure one can construct first the solutions of the eigenvalue problem of the Schr\"odinger
equation on RBRS in the form of so called stationary pre-solutions. These are the solutions which can be discontinues by themselves or by their first
derivatives on side traces on RBRS of the reflected RB. Having built the pre-solutions one can get the respective solutions for RB by a simple algebra
over the pre-solutions on RBRS. As a byproduct of such a procedure we get the result that energy spectra in RB are determined totally by periods of RBRS.

However it should be stressed that the pre-solutions are not semiclassical as well as the solutions to the Schr\"odinger equation in RB obtained by them. The unique junction with the
semiclassical constructions done in our earlier papers is made by the constructions of EPP and RBRS present in both the approaches. However the limitations of
possible boundary conditions which can be put on in the eigen problems to be solved are the same in both the approaches.

The paper is organized as follows.

In Sec.2 it is shown that any SS to SE in RB satisfying the Dirichlet or the Neumann boundary conditions on different sides of RB can be smoothly continued
on the whole EPP and further by periodic operations on the whole RBRS.

In Sec.3 the stationary pre-solutions to SE are constructed on EPP and RBRS.

In Sec.4 it is shown how SPS can be quantized on EPP and on RBRS according to the boundary conditions put on it on different sides of RB and
it is shown also how SS in RB is constructed by SPS.

In Sec.5 RBRS are built for several POCDRB - for the rectangular billiards, the equilateral triangle one, the rhombus-like billiards, the L-shape billiards, the
Bogomolny-Schmit billiards and the rectangular billiards with rectangular holes.

In Sec.6 SPS are constructed on RBRS built in Sec.5 by the Fourier series expansions the latter being the main tool of these constructions. The respective
quantization conditions are written and the role of the POC energy spectra in the quantizations is shown.

In Sec.7 the existence of SSS among of SPS is discussed and their positions in the quantization of POCDRB is established.

In Sec.8 the result of the paper are summarised and discussed.

The paper is completed by Appendices A-D.

\section{Continuation of SS into RBRS}

\hskip+2em Consider Fig.1 on which a side {\bf s} of RB {\bf A} is shown together with its image ${\bf B}$ done by this side. Consider {\bf A} at the
moment
as the original one while {\bf B} as its image in EPP generated by ${\bf A}$. Let $\Psi_A(x,y)$ be a stationary solution in {\bf A} satisfying some
boundary conditions on the side {\bf s}, i.e. the Dirichlet or the Neumann ones. A continuation $\Psi_B(x,y)$ of $\Psi_A(x,y)$ into {\bf B} can be done as
follows

\be
\Psi_B(-x,y)=-\Psi_A(x,y)
\label{1}
\ee
for the Dirichlet condition on the side {\bf s} and
\be
\Psi_B(-x,y)=\Psi_A(x,y)
\label{2}
\ee
for the Neumann one.

Making the above prescriptions of continuations of $\Psi_A(x,y)$ for all the other sides of {\bf A} we can continue $\Psi_A(x,y)$ into all the closest images
of {\bf A}. Next this procedure can be applied to images themselves to continue $\Psi_A(x,y)$ on the whole EPP.

However it must be noticed that not every
set of boundary conditions put on $\Psi_A(x,y)$ can allow its self-consistent continuation into EPP just described, i.e. EPP corresponding to a given RB selects in fact only some
limited number of such sets \cite{2}.

Suppose that ${\bf A}_i,\;i=1,...,2C$, with ${\bf A}_1={\bf A}$
are all images of {\bf A} in EPP. Then the set of allowed boundary condition in a given EPP can be characterized as follows
\begin{itemize}
\item prescribe to each image ${\bf A}_i$ of {\bf A} in its EPP a sign $\eta_i,\;\eta_i=\pm 1,\;i=1,...,2C$;
\item each side of ${\bf A}$ in EPP (there are $C$ images of a single side of {\bf A} in EPP) being common for two images ${\bf A}_i,\;{\bf A}_j$
gets then the sign $\eta_i\eta_j$;
\item the set $\Gamma=\{\eta_i:\eta_i=\pm 1,\;i=1,...,2C\}$ is called compatible if $\eta_i\eta_j$ for all $C$ images of each side of {\bf A} in EPP
is the same;
\item if the set $\Gamma$ is compatible then on the sides with $\eta_i\eta_j=+1$ the Neumann boundary conditions can be put on SPS while in the
opposite case - the Dirichlet ones.
\end {itemize}

Having continued $\Psi_A(x,y)$ on the whole EPP one can proceed its continuation on the whole RBRS just using periods defined by EPP and mentioned in Introduction.
In this way the continued $\Psi_A(x,y)$ becomes a stationary solution $\Psi(x,y)$ defined on the whole RBRS.

Suppose therefore that $P_i\in{\bf A}_i ,\;i=1,...,2C$, with $P_1\equiv(x,y)\in {\bf A}$ is a set of all images in EPP of the point $P_1$. Then we have the following identities
\be
\Psi_A(x,y)=\eta_i\Psi(P_i)=\frac{1}{2C}\sum_{j=1}^{2C}\eta_j\Psi(P_j)\nn\\
i=1,...,2C
\label{3}
\ee
where $\eta_1=1$

\begin{figure}
\begin{center}
\psfig{figure=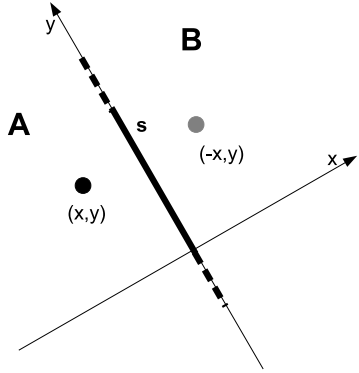,width=5 cm}
\caption{{\bf A} and {\bf B} - some two billiards images of EPP being the mirror reflections of each other in their common side {\bf s}}
\end{center}
\end{figure}

The wave function $\Psi(x,y)$ obtained in the above way has the following properties
\begin{enumerate}
\item it satisfies the Schr\"odinger equation in each point of RBRS;
\item it is periodic on RBRS with periods of RBRS;
\item it satisfies the Dirichlet or the Neumann conditions on the segments of RBRS which coincide with images of the sides of RB the emages of which cover
RBRS totally
\end{enumerate}

In the next section we will invert the above procedure of getting $\Psi(x,y)$ by finding so called pre-solutions to the stationary Schr\"odinger equation
on RBRS, i.e. not appealing to the solutions $\Psi_A(x,y)$.

\section{The stationary pre-solutions to the Schr\"odinger equation defined on RBRS}

\hskip+2em Let us now construct stationary pre-solutions to the Schr\"odinger equation on RBRS for a given RB, i.e. the solutions $\Psi(x,y)$ to the equation
\be
\frac{\p^2\Psi(x,y)}{\p x^2}+\frac{\p^2\Psi(x,y)}{\p y^2}+k^2\Psi(x,y)=0
\label{4}
\ee
deprived of some properties of continuity on RBRS. Here $k^2=2E$ and $E$ is energy of the system.

The stationary pre-solutions we are looking for are the stationary solutions to the\linebreak Schr\"odinger equation valid in the whole RBRS except
the traces of sides of RB on RBRS.

To this aim consider a rational billiards {\bf A} and EPP corresponding to it. Let ${\bf A}_i,\;i=1,...,2C$ be all the different images of {\bf A} with
${\bf A}_1\equiv {\bf A}$.
Then inside each image ${\bf A}_i$ and on its boundaries any SS $\Psi_{{\bf A}_i}(x,y;k)$ to the equation \mref{4} must have the form (see App.A)
\be
\Psi_{{\bf A}_i}(x,y;k)=\int_0^{2\pi}A_i(\phi)(\cos(kx\cos\phi)\cos(ky\sin\phi)+\cos(kx\cos\phi)\sin(ky\sin\phi)+\nn\\
                             \sin(kx\cos\phi)\cos(ky\sin\phi)+\sin(kx\cos\phi)\sin(ky\sin\phi))d\phi=\nn\\
          \int_0^{2\pi}{\tilde A}_i(\phi)e^{ik(x\cos\phi+y\sin\phi)}d\phi\nn\\
          i=1,...,2C
\label{5}
\ee
where $A_i(\phi)$ and ${\tilde A}_i(\phi)$ can be expanded both into the following Fourier series
\be
A_i(\phi)=A_{i0}+\sum_{p\geq 1}\ll(A_{i,p}^{(1)}\sin(p\phi)+A_{i,p}^{(2)}\cos(p\phi)\r)\nn\\
{\tilde A}_i(\phi)=A_{i,0}+\sum_{p\geq 1}\ll(-A_{i,2p}^{(1)}\sin(2p\phi)-iA_{i,2p-1}^{(1)}\sin((2p-1)\phi)+\r.\nn\\
                                     \ll. A_{i,2p}^{(2)}\cos(2p\phi)-iA_{i,2p-1}^{(2)}\cos((2p-1)\phi)\r)
\label{6}
\ee
i.e. in every RB the stationary $\Psi_{{\bf A}_i}(x,y;k)$ is always an interference of plane wave functions with different momenta but with the same energy.

Let us note further that according to \mref{5}-\mref{6} and App.A in each image ${\bf A}_i$ the solution $\Psi_{{\bf A}_i}(x,y;k)$ can be represented by the
following series
\be
\Psi_{{\bf A}_i}(x,y;k)=\nn\\
2\pi\ll(A_{i,0}J_0(k(x-x_i))J_0(k(y-y_i))+J_0(k(x-x_i))\sum_{r\geq 1}J_{2r-1}(k(y-y_i))A_{i,2r-1}^{(1)}+\r.\nn\\
                   \sum_{s\geq r\geq 1}(-1)^rJ_{2r}(k(x-x_i))J_{2s+1}(k(y-y_i))\ll(A_{i,2r+2s+1}^{(1)}+A_{i,2s-2r+1}^{(1)}\r)+\nn\\
                   \sum_{r>s\geq 0}(-1)^rJ_{2r}(k(x-x_i))J_{2s+1}(k(y-y_i))\ll(A_{i,2r+2s+1}^{(1)}-A_{i,2r-2s-1}^{(1)}\r)-\nn\\
                   \sum_{r>s\geq 0}(-1)^rJ_{2r+1}(k(x-x_i))J_{2s+1}(k(y-y_i))\ll(A_{i,2r+2s+2}^{(1)}-A_{i,2r-2s}^{(1)}\r)-\nn\\
                   \sum_{s>r\geq 0}(-1)^rJ_{2r+1}(k(x-x_i))J_{2s+1}(k(y-y_i))\ll(A_{i,2r+2s+2}^{(1)}+A_{i,2s-2r}^{(1)}\r)-\nn\\
                  \sum_{s\geq 0}(-1)^sJ_{2s+1}(k(x-x_i))J_{2s+1}(k(y-y_i))A_{i,4s+2}^{(1)}+\nn\\
                    \sum_{r\geq 1}\ll(J_0(k(x-x_i))J_{2r}(ky)+(-1)^rJ_0(k(y-y_i))J_{2r}(kx)\r)A_{i,2r}^{(2)}+\nn\\
                   \sum_{r,s\geq 1}(-1)^rJ_{2r}(k(x-x_i))J_{2s}(k(y-y_i))\ll(A_{i,2r+2s}^{(2)}+A_{i,2|r-s|}^{(2)}\r)-\nn\\
                   J_0(k(y-y_i))\sum_{r\geq 0}(-1)^rJ_{2r+1}(k(x-x_i))A_{i,2r+1}^{(2)}+\nn\\
               \ll.\sum_{r\geq 0,s\geq 1}(-1)^rJ_{2r+1}(k(x-x_i))J_{2s}(k(y-y_i))\ll(A_{i,2r+2s+1}^{(2)}+A_{i,|2r-2s+1|}^{(2)}\r)\r)\nn\\
         (x,y),\;(x_i,y_i)\in P_i,\;\;\; i=1,...,2C
\label{9}
\ee
where $J_n(z),\;n\geq 0$, are the Bessel functions. Of course in each image ${\bf A}_i$ the coefficients $A_{i,p}^{(j)}$ of the series can be different
and it is also assumed that each image ${\bf A}_i$ is contained in the area $S_i$ of convergence of the series \mref{9}, see App.A.

Note that there are no any relation between the solutions $\Psi_{{\bf A}_i}(x,y;k)$ except the value of the wave number $k$ which is the same for each of them.

The pre-solution $\Psi(x,y;k)$ can be now defined in every point $P\in$ EPP by
\be
\Psi(P)=\Psi_{{\bf A}_i}(P),\;\;\;P\in {\bf A}_i
\label{7}
\ee

The solution $\Psi(x,y;k)$ can be then distributed on the whole RBRS by shifting repeatedly EPP by its periods.

$\Psi(x,y;k)$ defined in this way on RBRS has the following basic properties
\begin{itemize}
\item it is a solution to SE inside each image of {\bf A} in RBRS;
\item it can be discontinuous on the side traces of ${\bf A}$ in EPP and in RBRS together with its first derivatives;
\item it is periodic on RBRS with the periods of RBRS;
\end{itemize}

\section{Solving the stationary state problem for RB}

\hskip+2em In principle one can then use the form \mref{9} to finish the construction of $\Psi(x,y;k)$ by
\begin{itemize}
\item demanding it to satisfy the condition of periodicity on each pair of the boundary sides of EPP being distant with respect to each
other by a period or simply demanding $\Psi(x,y;k)$ to be periodic on RBRS;
\item matching the representations \mref{9} of $\Psi(x,y;k)$ on the common sides of the images in which they are defined according to
\begin{itemize}
\item $\Psi(x,y;k)$ must be continuous on all images of a side of {\bf A} in EPP or in RBRS if the Dirichlet boundary conditions of $\Psi_{\bf A}(x,y;k)$ are to be
satisfied on the side;
\item the first derivatives of $\Psi(x,y;k)$ must be continuous on all images of a side of {\bf A} in EPP or in RBRS if the Neumann boundary conditions are
demanded for $\Psi_{\bf A}(x,y)$ on the side.
\end{itemize}
\end {itemize}

The above conditions would then provide us with a system of linear homogeneous equations
determining both the coefficients $A_{i,p}^{(i)}(k), i=1,2,\;p\geq 0$, in\mref{9} and the corresponding energy spectra $k_n,\;n\geq 0$ of
$\Psi(x,y;k_n)$ the latter being a solution to the vanishing determinant of the system.

If $\Psi(x,y;k_n)$ is determined by the above procedure then SS $\Psi_{\bf A}(x,y;k_n)$ is now obtained by substituting $\Psi(x,y;k_n)$ to the sum on the right hand side
of \mref{3}, i.e. we have
\be
\Psi_{\bf A}(x,y;k_n)=\frac{1}{2C}\sum_{i=1}^{2C}\eta_i\Psi(x_i,y_i;k_n),\;\;\;\;n\geq 0
\label{8}
\ee
where $(x_i,y_i),\;i=1,...2C$, are all the mirror images of the point $(x,y)=(x_1,y_1)$ in EPP and $\eta_1=+1$.

It is then easy to check that $\Psi_{\bf A}(x,y;k_n),\;n\geq 0$ are then the respective solutions satisfying on the
boundary of {\bf A} the conditions defined by the chosen coefficients $\eta_i$ in \mref{8}. Since the
boundary conditions define the corresponding energy spectrum of $\Psi_{\bf A}(x,y;k_n),\;n\geq 0$ uniquely we conclude that the latter must be a part of the spectrum of
$\Psi(x,y;k_n)$ itself.

It should be obvious that the formula \mref{8} containing the pre-solution $\Psi(x,y;k_n)$ defines also a continuation of $\Psi_{\bf A}(x,y;k_n)$ on both the whole EPP and
the whole RBRS. Namely, if the point $P_i=(x_i,y_i;k_n)$ is the image of the point $P=(x,y;k_n)\in{\bf A}$ in the image ${\bf A}_i$ of {\bf A} in EPP
then we have
\be
\Psi_{{\bf A}_i}(x_i,y_i;k_n)=\eta_i\Psi_{\bf A}(x,y;k_n)=\frac{1}{2C}\sum_{j=1}^{2C}\eta_i\eta_j\Psi(x_j,y_j;k_n)
\label{8b}
\ee
where $\Psi_{{\bf A}_i}(x_i,y_i;k_n),\;i=2,...,2C$ are created from $\Psi_{\bf A}(x,y;k_n)$ by the rules \mref{1}-\mref{2}.

Therefore solving the stationary problem in RB in the above way we get as one of its general results the following
\begin{itemize}
\item the respective wave functions and energy spectra of any RB are defined completely by the periodic structure of EPP corresponding to it.
\end{itemize}

While the procedure described above and the conclusion which follows from it have the property of being general the form \mref{9} of $\Psi(x,y;k)$
does not allow us for other valuable conclusions. Such an offer however can be got by confining the class of the considered RB to the POCD ones.
We shall do it in the next sections considering several examples of such billiards.

\section{RBRS for simple POCDRB}

\hskip+2em To get some feelings how RBRS can be constructed and used we shall consider several examples of their constructions starting from the two
integrable cases of the rectangular billiards and the equilateral triangle one considering farther the rhombus billiards, the L-shape billiards with
rational and irrational relations between their parallel sides, the triangular billiards of Bogomolny and Schmit \cite{6} and the two rectangular billiards
with a rectangular hole inside it - one with the sides of the inner rectangle parallel to the outer one and a second rotated by $\pi/4$ with respect
to the "mother" rectangle.

Nevertheless there is a standard way of construction of RBRS exploring POCs which goes along the following steps
\begin{enumerate}
\item fixing all possible branchpoints of EPP, i.e. all its vertices with the angles $\pi p/q$ where $p>1$;
\item choosing a period of EPP (which will be assumed to be "vertical") the direction of which excludes AOC;
\item cutting EPP along all singular diagonals parallel to the chosen period into finite stripes the non-diagonal boundaries of which are pieces of the
boundaries of EPP but avoiding sides which are common for two images of RB in EPP;
\item extending each such a finite stripe in the direction of the period by gluing it with others along its non-diagonal boundaries which are joined
by periods in EPP - the glued pieces are identical being two copies of the same side element of RB;
\item continuing such a gluing by exhausting all pieces of the cut EPP allowing for that - the maximal "vertical" stripe obtained in this way is
bounded from up and down by two copies of the same piece of a side element of RB;
\item forming a vertical POC of the identical copies of the maximal strip by gluing repeatedly its subsequent copies, i.e. the period of such a POC
is equal to the height of the strip which it is formed of. The vertical boundaries of both the sides of POC got in this way are built periodically
of the sides of RB or of the segments linking two vertices of RB with $p>1$;
\item forming a sheet by gluing a single or several POCs with themselves along their vertical boundaries - such a gluing must satisfy the following
rules
\begin{itemize}
\item two POC are glued along the sides of RB exclusively;
\item glued sides of RB must be the same;
\item the glued vertices must be the images of the same vertex of RB and the segments linking them must fit exactly to each others on both the
glued POC;
\item the fitted segments of two POCs form then two edges of the cut of the sheet;
\item two pieces of RB glued along its side or being in the immediate neighbourhood of each other separated by a cut must be mirror images of each
other;
\end{itemize}
\item gluing the obtained sheets between themselves by the respective cuts which must fit to each other by their construction to get connected RBRS.
\end{enumerate}

It should be stressed that a form of RBRS depends on a choice of period used in constructing the respective POCs.

The above rules will be applied in the examples considered below.

Let us start with the rectangular billiards and the equilateral triangle one shown in Fig.2A,D respectively. Their corresponding EPP are shown in
Fig.2B,E together with the two independent periods
${\bf D}_1$ and ${\bf D}_2$ while their RBRS are shown in Fig.2C,F clearly exposing their periodic structure on the plane which two periods ${\bf D}_1$
and ${\bf D}_2$ act on, see Fig.2D. Distributions of the mirror images of the point $(x,y)$ of Fig.2A are also shown on all the figures.

The rhombus billiards with the smaller angles equal to $\pi/3$ is shown on Fig.3A. The description under the figure explains its content. This case
was used also to demonstrate a dependence of the sheet structure of RBRS on the choice of the period direction.

\begin{figure}
\begin{center}
\psfig{figure=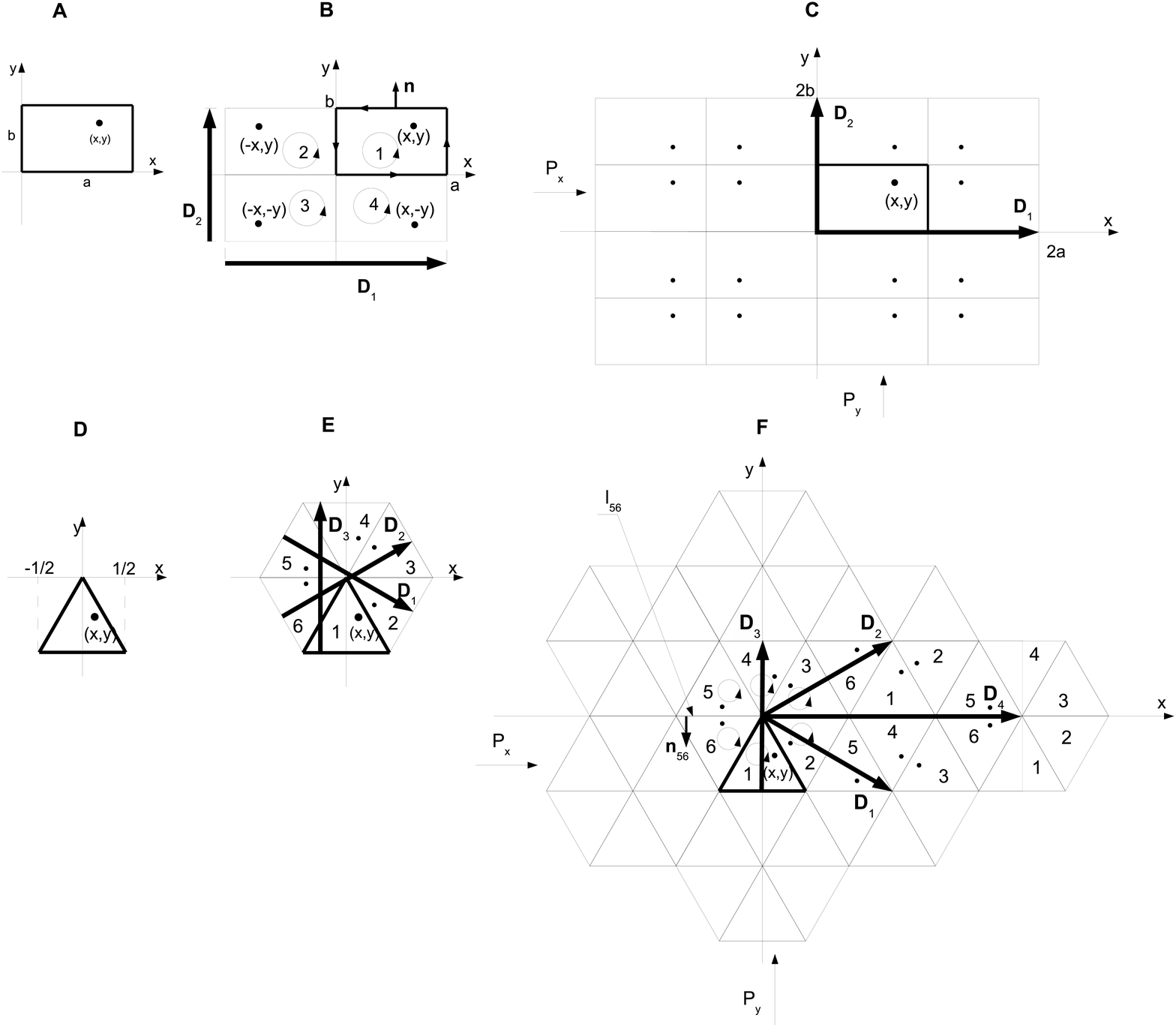,width=15 cm}
\caption{{\bf A,D} - the rectangular and  the equilateral billiards respectively, {\bf B,E} - their respective EPP with two periods ${\bf D}_1$ and
${\bf D}_2$, {\bf C,F} - the planes being the corresponding RBRS generated
periodically by the respective EPP using the periods ${\bf D}_1$ and ${\bf D}_2$}
\end{center}
\end{figure}
\begin{figure}
\begin{center}
\psfig{figure=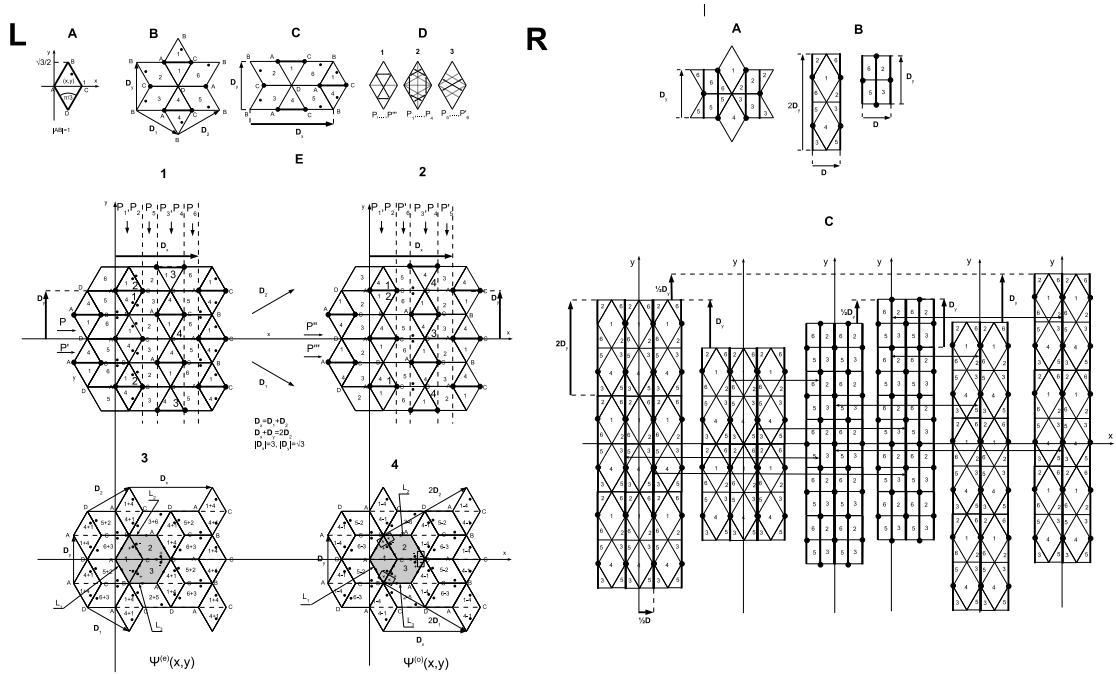,width=15 cm}
\caption{Left ({\bf L}): {\bf A} - the $\pi/3$-rhombus-like billiards; {\bf B} - its respective EPP with its periods ${\bf D}_i,\;i=1,2,y$, cut horizontally
parallel to the period ${\bf D}_x={\bf D}_1+{\bf D}_2$; {\bf C} - the element which the sheets {\bf E}1 and {\bf E}2
of RBRS are periodically formed of; {\bf D}1-{\bf D}3 - the respective POCs in the rhombus shown on
the sheets {\bf E}1 and {\bf E}2 of the corresponding RBRS; {\bf E}3-{\bf E}4 - the $x,y$-planes which the components $\Psi^{(e)}(x,y)$ and $\Psi^{(o)}(x,y)$
of SPS $\Psi(x,y)$ are defined on. Right ({\bf R}):
the alternative vertical cutting of EPP, i.e. parallel to the period ${\bf D}_y$ leading us to a different sheet structure of RBRS which is glued in such a
case of the six sheets shown. Arrows show the way of gluing the sheets between themselves.}
\end{center}
\end{figure}
\begin{figure}
\begin{center}
\psfig{figure=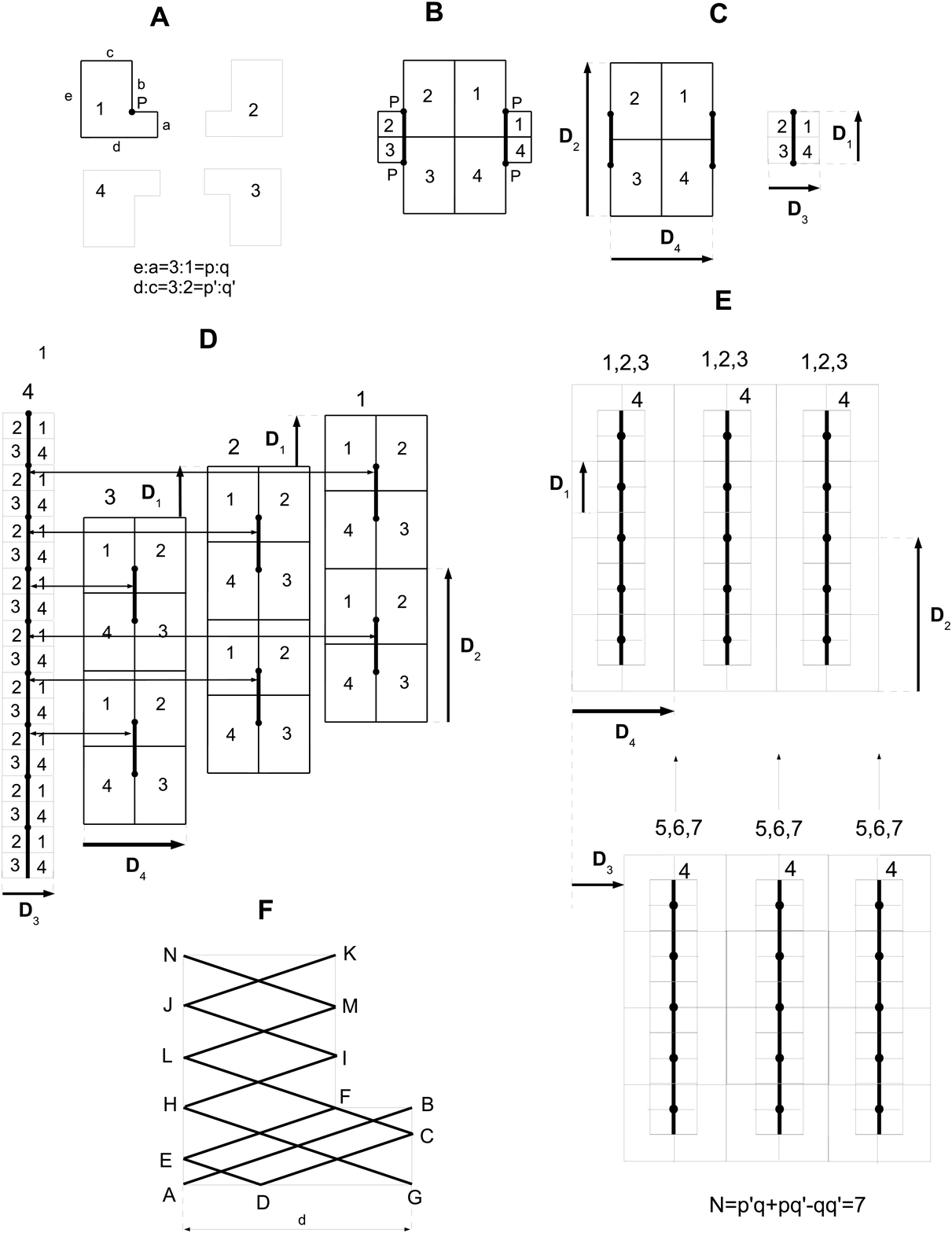,width=13 cm}
\caption{{\bf A} - the L-shape billiards with rational relations between its sides (the doubly rational billiards); {\bf B} - its respective EPP
cut between the branch points defined by the point $P$; {\bf C} - the two elements glued of the two maximal stripes each defining two types of
sheets of RBRS with the two pairs ${\bf D}_1,\;{\bf D}_2$ and ${\bf D}_3,\;{\bf D}_4$ of the respective periods acting on the sheets;
{\bf D}, {\bf E} - the successive steps of gluing the elements of {\bf C}
to get RBRS with the seven sheets. $N=p'q+pq'-qq'$ is a general formula for a number of sheets for L-shape billiards with rational relations between
their sides; {\bf F} - the magnified L-shape billiards {\bf A} with the four singular diagonals AB, CDEF, GHIJK, FLMN of the four POCs parallel to
each other on RBRS if the side $d$ is rational. If $d$ is irrational the two POCs with the common diagonal GHIJK become AOCs.}
\end{center}
\end{figure}
\begin{figure}
\begin{center}
\psfig{figure=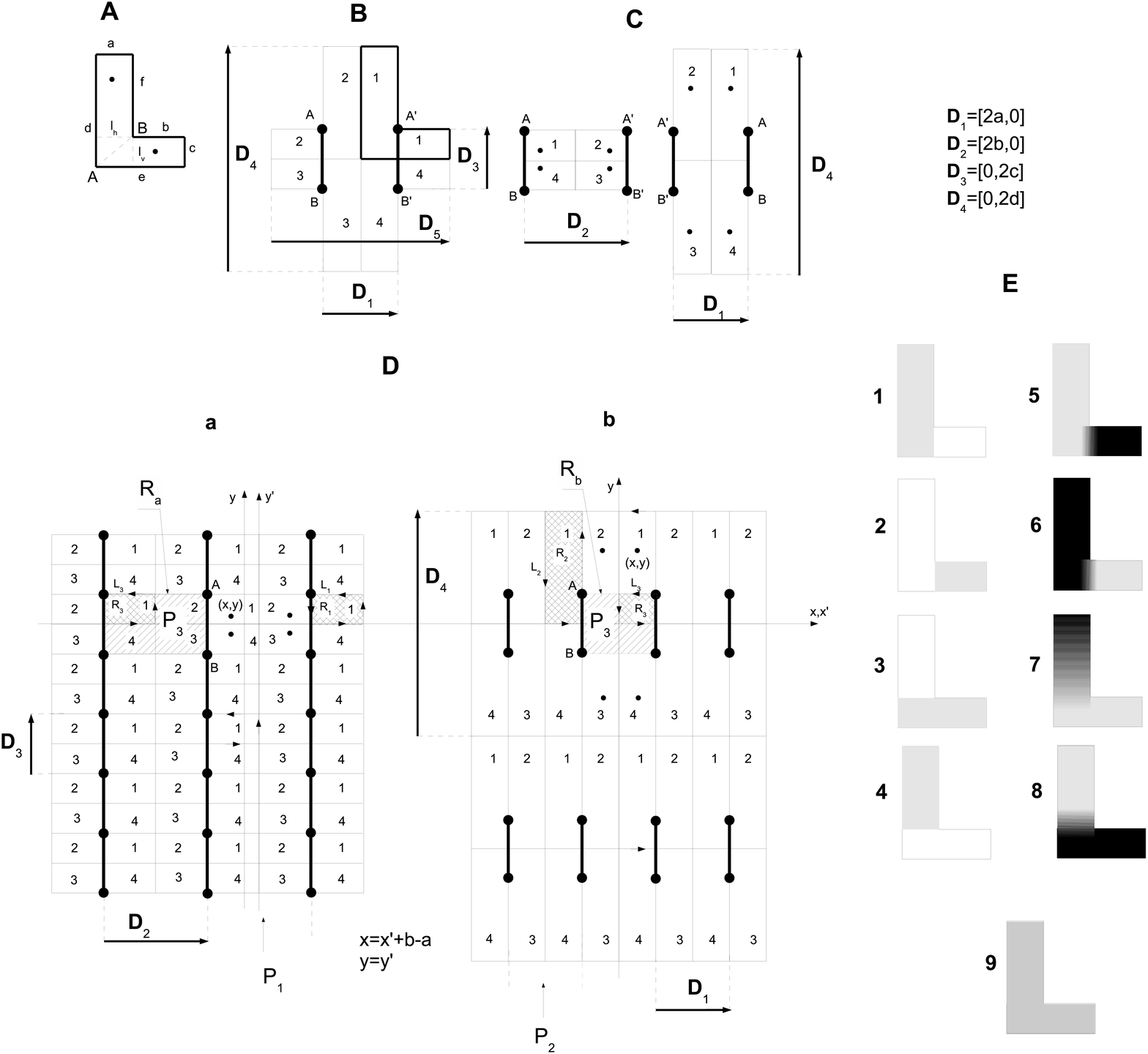,width=12 cm}
\caption{{\bf A} - the L-shape billiards with irrational relation between its parallel sides; {\bf B} - its respective EPP
with the periods ${\bf D}_1,\;{\bf D}_2,\;{\bf D}_3$ and ${\bf D}_5$; {\bf C} - two elements which sheets of RBRS are built of; {\bf D} - two kinds of infinitely
many sheets of RBRS. Each sheet of the type {\bf a} is glued with a sheet of the type {\bf b} by a single cut (A-B in the figure) and vice versa;
The $x,y$-axes coincide on all the sheets; {\bf E} - 1-4 - the hypothetical superscar states which cannot be realized in the billiards, 5-8 - their
resonant realizations in the billiards and 9 - a possible superscar state in the doubly rational L-shape billiards.}
\end{center}
\end{figure}
\begin{figure}
\begin{center}
\psfig{figure=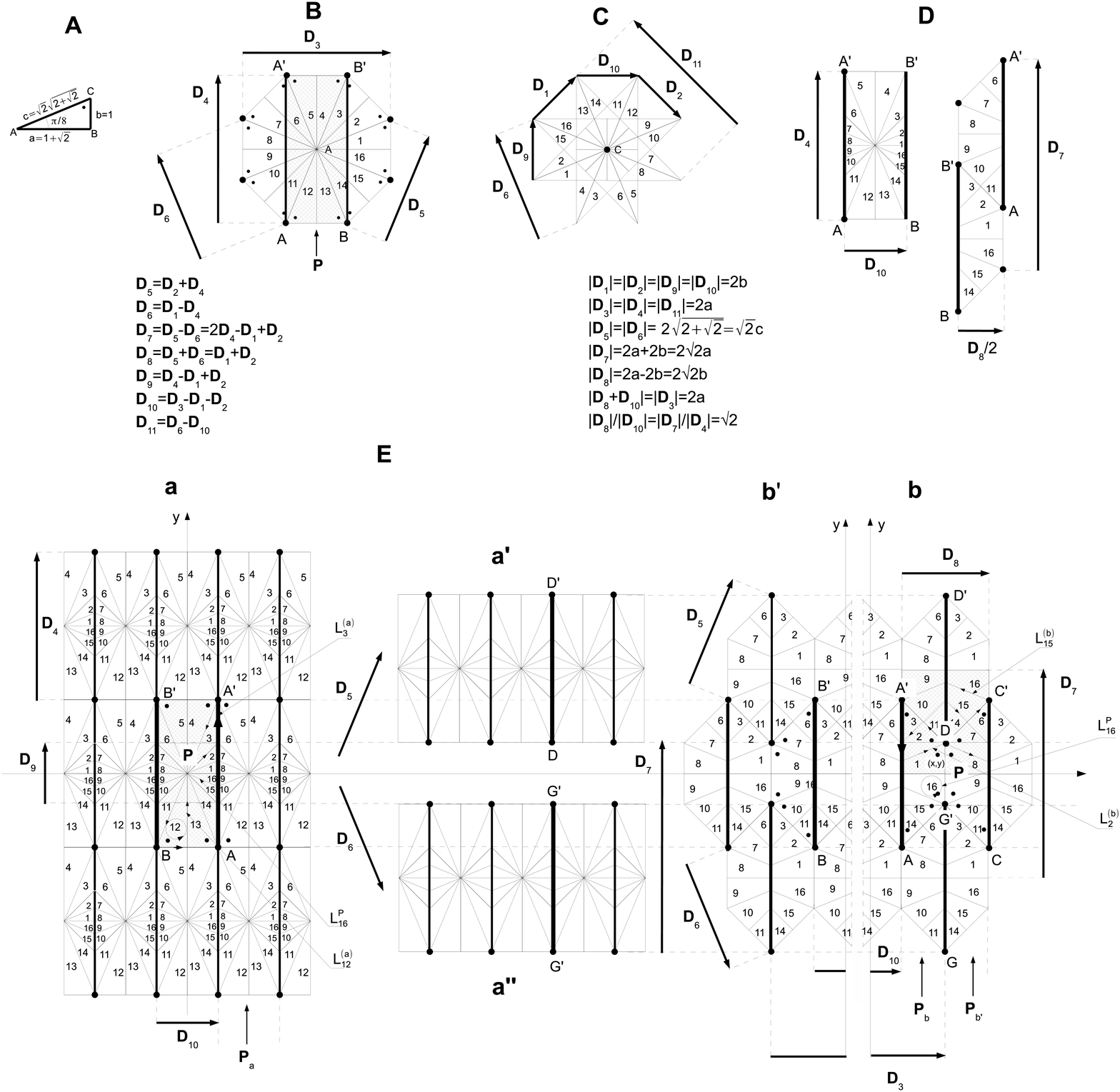,width=15 cm}
\caption{{\bf A} - the Bogomolny-Schmit billiards with $g=2$; {\bf B} - its respective EPP
with the periods ${\bf D}_3,\;{\bf D}_4$ shown and with the cuts $A-A'$ and $B-B'$ used to form the respective RBRS and with POC {\bf P} marked; {\bf C} - another form of EPP with
the periods ${\bf D}_1,\;{\bf D}_2$ collecting with the previous ones
the set of the four independent periods; {\bf D} - the two maximal stripes defining POCs which the sheets {\bf a} and {\bf b} of RBRS are built of;
{\bf E} - two types of the RBRS
sheets - a sheet of the type {\bf a} is glued only with a single one of the type {\bf b} and vice versa. Period relations between the sheets and
POCs ${\bf P}_a,\;{\bf P}_b,\;{\bf P}_{b'}$ are also shown in the figure. RBRS is now built of infinitely many sheets of the type
{\bf a} and of the type {\bf b}. The $x,y$-axes coincide on all the sheets.}
\end{center}
\end{figure}
\begin{figure}
\begin{center}
\psfig{figure=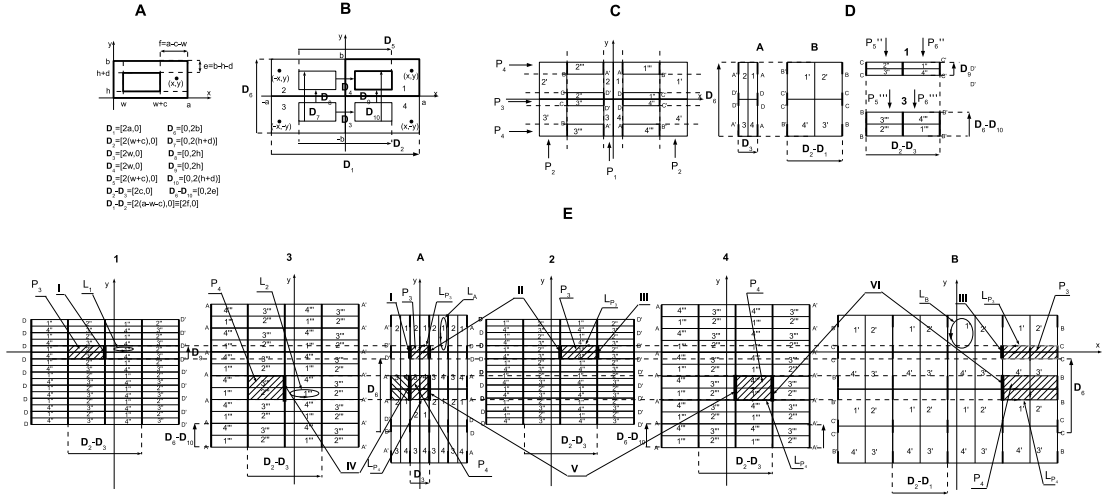,width=15 cm}
\caption{{\bf A} - the rectangular billiards with the rectangular hole corresponding to the multitorus with $g=5$; {\bf B} - its respective EPP;
{\bf C} - EPP cut respectively with POCs $P_3,P_3',P_4,P_4'$; {\bf D} - the
four elements of EPP which the sheets of RBRS is glued of; {\bf E} - the six sheets of RBRS selected from infinitely many of them glued between
themselves by the cuts $I,...,VI$ - each of the sheet ${\bf 1},\;{\bf 2},\;{\bf 3},\;{\bf 4}$ is glued with each sheet of
the type {\bf A} or {\bf B} by a single cut only and vice versa, i.e. there is no gluing between the sheets of the same type. The $x,y$-axis
are the same for all the sheets.}
\end{center}
\end{figure}
\begin{figure}
\begin{center}
\psfig{figure=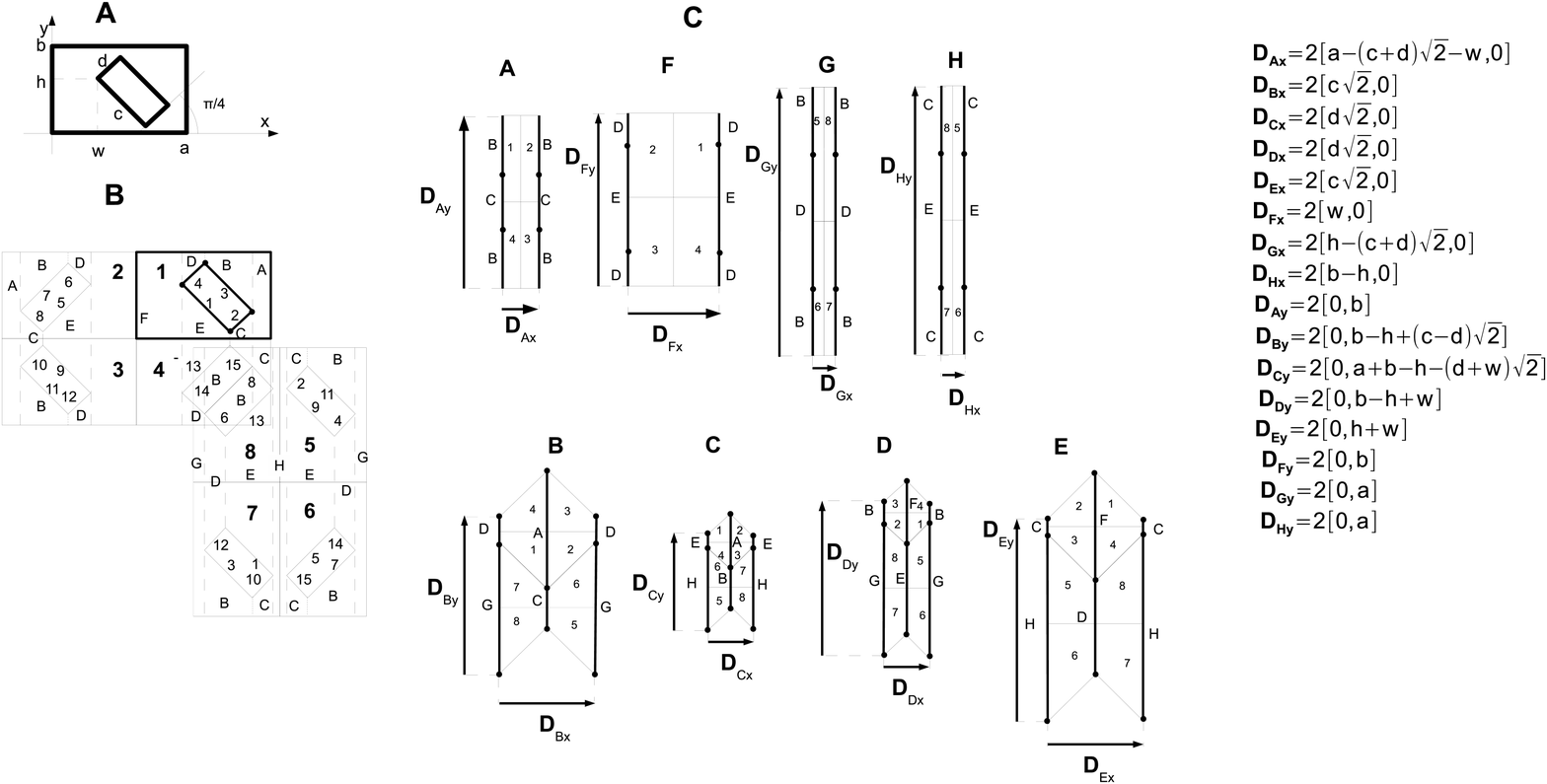,width=15 cm}
\caption{{\bf A} - the rectangular billiards with the rectangular hole rotated by $\pi/4$ with respect to the outer rectangle corresponding to the
multitorus with $g=9$ - its respective sizes are assumed to be incommensurable; {\bf B} - its respective EPP with its all vertical POCs $A,...,H$
and with their singular diagonals crossing the branch points of RBRS;
{\bf C} - EPP cut respectively along POCs and glued along their parallel boundary sides providing eight elements {\bf A},...,{\bf H} each of which
distributed on its own plane according to the periods
attributed to it defines a sheet of the corresponding RBRS with the same name. These eight different sheets are next glued with each other along
the cuts $A,...,H$, marked on the elements so that a cut $I,\;I=A,...,H,$ on a sheet ${\bf J},\;{\bf J}={\bf A},...,{\bf H}$, is glued with the
cut $J$ on the sheet {\bf I} - two different sheets are glued by a single cut only because of assumed incommensurability of the billiards sides.
There are infinitely many sheets building this RBRS. However due to the periodicity of RBRS the eight mentioned sheets with some finite number of
their copies are enough for constructing the corresponding SPS.}
\end{center}
\end{figure}

Fig.4 shows the L-shape billiards with the rational relations between its sides. The respective steps in constructions of its EPP and RBRS are shown in
the parts {\bf C}-{\bf E} of the figure. The number of sheets of RBRS for the case on the figure is equal to seven. However for a general L-shape billiards
with the rational relations between its sides defined by $e:a=p:q$ and $d:c=p':q'$ where the integers in the pairs $p,q$ and $p',q'$ are coprime there
is a general results for the number $N$ of sheets of the corresponding RBRS, namely $N=pq'+p'q-qq'$. Fig.4F shows the L-shape billiards considered
magnified in which the singular diagonals for the four POCs are sketched. The POCs are as such if the size $d$ of the billiards is rational but two of
them are transformed into the corresponding AOCs if $d$ becomes irrational.

Fig.5 shows the the example of RBRS with an infinite number of sheets provided by the L-shape billiards with the irrational relation between its vertical
and horizontal sides. The construction of the corresponding RBRS is selfexplaining on the figure.

Fig.6 shows RBRS constructed for the famous Bogomolny-Schmit billiards \cite{6}. The surface is built of infinitely many sheets in this case.

A billiards with a hole inside it is represented in Fig.7 by the rectangular billiards with the rectangular hole the sides of which are parallel to the
"mother" rectangle but the relations between the vertical as well the horizontal sides of the inner and outer rectangles are assumed to be irrational.
Then the number of sheets of RBRS for the case is infinite.

As the last example of the composition of RBRS for the rational billiards is presented the rectangular one with the rotated rectangular hole, see
Fig.8. It is clear that the more complicated billiards the more complicated becomes the respective RBRS constructed for it.

Obviously each RBRS is invariant on actions of any of its periods. Note however that a period of RBRS can act on the latter in the following two
different ways  -
it can transform a sheet of RBRS into itself or it can transform one sheet of RBRS into another. Both the actions are seen on the figures 1-6 and are
directly related
to the way which RBRS was constructed by. In the case of Fig.2 since the respective RBRS is built of a single sheet each period transforms the sheet
and the corresponding RBRS into itself while on the cases of Fig.Fig.3-8 the constructions of
the corresponding RBRS are possible only by shifting some glued sheets with respect to some other by some periods of RBRS. Therefore in the first case
the points of RBRS joined by a period lie on the same sheet of RBRS while in the second one the respective points can lie on two different sheets. For a
function defined on RBRS and respecting its periodic structure it means that in the first case it can be periodic on a single sheet while
in the second case its periodicity can be realized only if all sheets of RBRS are considered simultaneously.

\section{SPS and SS in POCDRB}

\hskip+2em For POCDRB for which their respective RBRS are periodically constructed of sheets which structure is also periodic the proper tool for
building $\Psi(x,y;k)$ is representing them by the
respective Fourier series. The latter representations will be used therefore notoriously in the examples considered
in the next subsections where for RB shown in Fig.Fig.2-7 the respective SPS $\Psi(x,y;k)$ will be constructed showing explicitely how the periods
of EPP can form the solutions to SE.

On the other hand the domain of the plane in which the Fourier series with the periods {\bf a} and {\bf b} is defined has always the form of
the parallelograms which sides are made of the periods of the series. In such a domain $\Psi(x,y;k)$ is reconstructed by the series only when
$\Psi(x,y;k)$ itself or their
derivatives are continuous in this domain, i.e. the series do not reconstruct $\Psi(x,y;k)$ in vicinities of points of the domain in which
$\Psi(x,y;k)$ or its
derivatives are identified by the series as being discontinuous (Gibbs effect). If such irregularities of $\Psi(x,y;k)$ are real then their existence
is visible in the structure of the Fourier series coefficients corresponding to the derivatives of $\Psi(x,y;k)$. In other cases, i.e. for points
lying on the boundary of the period parallelogram the respective discontinuities of expanded $\Psi(x,y;k)$ are frequently apparent appearing as a
result of a form of the chosen parallelogram. A remedy for avoiding such discontinuities is making the Fourier expansion for $\Psi(x,y;k)$
in another parallelogram containing the singular boundaries of the previous one inside. However both the expansions need then to be fully identified
which in the case of $\Psi(x,y;k)$ means identifications of the respective series not only for $\Psi(x,y;k)$ itself but also for its derivatives to
take also into account the existing real discontinuities of $\Psi(x,y;k)$. The respective procedure will be applied to the cases of the billiards
considered in the followed subsections.

The first thing however necessary to develop the method is the construction of the respective RBRS with a definite periodic structure.
The respective description how to do it was given by the points 1. - 8. in Sec.2. It should be completed by steps describing the way of using RBRS to define
$\Psi(x,y;k)$ on it. They are the following
\begin{enumerate}
\setcounter{enumi}{8}
\item choosing a set of sheets which are periodically repeated in the structure of RBRS and are glued between themselves;
\item defining on each chosen sheet having some two independent periods a branch of SPS in the form of the Fourier series expansion (FSE);
\item matching the branches defined on different sheets using POCs which are passing by these different sheets;
\item forming the matching conditions up to the second derivatives including the Schr\"odinger equations.
\end{enumerate}

It will be assumed further that $\Psi(x,y;k)$ and their derivatives up to the second ones are continuous inside each image of EPP, i.e. the possible
discontinuities are localized only on the boundaries of the images.

The steps 1.-12. sketched above define a complete procedure to get SS and the corresponding energy spectrum for the quantized stationary motion in
RB.

It should be stressed here that despite the fact that the above procedure defines $\Psi(x,y;k)$ on the two dimensional surface which RBRS seems to be
its multi-sheeted
structure allows $\Psi(x,y;k)$ to be multi-periodic on it, i.e. on each sheet of RBRS there can be (at most) a pair of independent (in the space of integers)
periods of  $\Psi(x,y;k)$ but on different sheets such pairs can be different. The latter means also that a number of sheets of RBRS must be always no less
than the genus $g$ of the respective multi-torus.

Below the
procedure will be applied to the following examples of RB: the rectangular one, the equilateral triangle one, the rhombus-like one, the L-shape billiards,
Bogomolny-Schmit one and to the rectangular billiards with the rectangular holes.
The first eight of the point above are realized simply by making the respective pictures for the cases mentioned - Fig.Fig.2.-7.

However in order not to enlarge the volume of the paper excessively only the first four billiards will be considered in necessary details while the
remaining
ones only schematically assuming that the detailed applications of the method to the latter cases are clear due to the first four ones.

\subsection{SPS for the rectangular billiards}

\hskip+2em Considering this case seems to be a trivial effort but it offers instead a clear presentation of some elements of the respective procedure. Obviously
RBRS for the rectangular billiards is simply a plane with the two independent periods ${\bf D}_1,{\bf D}_2$ as it is shown in Fig.2C

According therefore to the general prescription given above to construct SPS in EPP given by Fig.2B we have to define in each image of it a stationary
solution to SE using for example the form \mref{9}. There are four such images $1,...,4$ and respectively we have four solutions $\Psi_i(x,y),\;i=1,...,4$,
independent of each other by a while. Nevertheless since the solutions are fixed in EPP they are also periodically fixed in corresponding RBRS allowing for
the Fourier series expansion of the respective SPS $\Psi(x,y;k)$ which has two independent periods ${\bf D}_x=[2a,0]$ and ${\bf D}_y=[0,2b]$ on RBRS. Therefore
we have
\be
\Psi^{(FS)}(x,y;k)=
\sum_{i,j=1,2}\sum_{m,n\geq 0}X_{ij,mn}f_i\ll(m\pi\frac{x}{a}\r)f_j\ll(\pi n\frac{y}{b}\r)\nn\\
f_1(x)=\sin x,\;f_2(x)=\cos x,\;f_i'(x)=(-1)^{i+1}f_{i+1}(x),\;f_{i\pm 2}(x)=f_i(x),\;i=1,2
\label{10}
\ee

Assuming none boundary condition to be satisfied by $\Psi(x,y;k)$ we have for its Fourier expansion \mref{10}
\be
\Psi^{(FS)}(0,y;k)=\ll\{\ba{lr}
                        \fr(\Psi_1(0,y;k)+\Psi_2(0,y;k)),&0<y<b\\
                        \fr(\Psi_3(0,y;k)+\Psi_4(0,y;k)),&-b<y<0
                        \ea\r.\nn\\
\Psi^{(FS)}(a,y;k)=\ll\{\ba{lr}
                        \fr(\Psi_1(a,y;k)+\Psi_2(-a,y;k)),&0<y<b\\
                        \fr(\Psi_3(-a,y;k)+\Psi_4(a,y;k)),&-b<y<0
                        \ea\r.
\label{11}
\ee
and similar relations on other sides of the rectangle.

Obviously, $X_{ij,mn}$ are defined by
\be
X_{ij,mn}=\frac{1}{ab}\int_{-a}^adx\int_{-b}^bdy\Psi(x,y;k)f_i\ll(\pi m\frac{x}{a}\r)f_j\ll(\pi n\frac{y}{b}\r)=\nn\\
\frac{4}{ab}\int_0^adx\int_0^bdy\Psi^{(ji)}(x,y;k)f_i\ll(\pi m\frac{x}{a}\r)f_j\ll(\pi n\frac{y}{b}\r)
\label{11e}
\ee
where
\be
\Psi^{(ij)}(x,y;k)=\nn\\
\frac{1}{4}(\Psi_1(x,y;k)+(-1)^j\Psi_2(-x,y;k)+(-1)^{i+j}\Psi_3(-x,-y;k)+(-1)^i\Psi_4(x,-y;k))\nn\\
0<x<a,\;0<y<b,\;\;i,j=1,2
\label{11a}
\ee

The latter have also the following extension into EPP
\be
\Psi_{ext}^{(ij)}(x,y;k)=\ll\{\ba{lr}
                               \Psi^{(ij)}(x,y;k),&0<x<a,\;0<y<b\\
                         (-1)^j\Psi^{(ij)}(-x,y;k),&-a<x<0,\;0<y<b\\
                         (-1)^i\Psi^{(ij)}(x,-y;k),&0<x<a,\;-b<y<0\\
                         (-1)^{i+j}\Psi^{(ij)}(-x,-y;k),&-a<x<0,\;-b<y<0
                         \ea\r.
\label{11b}
\ee
with the following Fourier expansion on the corresponding RBRS
\be
\Psi_{ext}^{(ij,FS)}(x,y;k)=\sum_{m,n\geq 0}X_{ji,mn}(k)f_j\ll(\pi m\frac{x}{a}\r)f_i\ll(\pi n\frac{y}{b}\r)\nn\\
-\infty <x,y<+\infty
\label{11c}
\ee

Moreover we have of course
\be
\Psi(x,y;k)=\sum_{i,j=1,2}\Psi_{ext}^{(ij)}(x,y;k)=\ll\{\ba{lr}
                               \Psi_1(x,y;k),&0<x<a,\;0<y<b\\
                         \Psi_2(x,y;k),&-a<x<0,\;0<y<b\\
                          \Psi_3(x,y;k),&-a<x<0,\;-b<y<0\\
                         \Psi_4(x,y;k),&0<x<a,\;-b<y<0
                         \ea\r.
\label{11d}
\ee

Let us now note that not every SS in the rectangular billiards can be constructed by the formula \mref{8}, i.e. the formula allows for the solutions
the boundary conditions for which coincide with the following four sequences of the signs $\eta_i,\;i=1,...,4$, in \mref{8} $(++++),(+--+),(++--),(+-+-)$.
Denote by $1$ the respective Dirichlet boundary condition put on a side of the rectangle and by $2$ - the Neumann one. Then
the sets of the signs mentioned correspond to the following pairs of the boundary conditions put on the pairs of the parallel sides $ab$ -
$22$, $21$, $12$ and $11$. If now the index $i$ in $\Psi^{(ij)}(x,y),\;i,j=1,2$, corresponds to the $a$-sides and
and the $j$ one to the parallel $b$-sides then $\Psi^{(ij)}(x,y),\;i,j=1,2$, represent the functions \mref{8} formed for the respective boundary
conditions mentioned. Let us note however that $\Psi^{(ij)}(x,y)$ themselves do not satisfy any of the conditions mentioned as long as
the respective conditions 1.- 2. of sec.5 are put on SPS itself.

It is worth to make also the following note
\begin{enumerate}
\item to construct SPS as a periodic function on RBRS it is not necessary to use the assumption that the functions $\Psi_i(x,y;k),\;i=1,...,4$ have to
satisfy the Schr\"odinger equation. In fact they can be arbitrary allowing only for the Fourier representation of $\Psi(x,y;k)$ on RBRS;

and necessary the following one
\item by using the Fourier series \mref{10} for SPS one can construct functions \mref{11a} the Fourier series \mref{11c} of which satisfy the boundary
conditions on the rectangle despite the fact that none of the respective conditions is put on SPS itself, i.e. $\Psi^{(ij)}(x,y),\;i,j=1,2$ as given
by \mref{11e} do not satisfy the respective conditions on the rectangle boundaries.
\end{enumerate}

The latter note is extremely important since it shows that the Fourier series cannot be applied in vicinities of points where functions represented
by the series are discontinuous including points distant by respective periods. However the latter discontinuities can be frequently of a "technical"
origin, i.e.
the points distant by a period can be regular for $\Psi(x,y;k)$ but can lie on the parallel boundaries of the Fourier expansion and the series treats
them as irregular if $\Psi(x,y;k)$ takes in these points different values. For such a case a remedy for it is using for $\Psi(x,y;k)$ another Fourier
representation in which the points lie inside the expansion domain. Of course it needs then matching up such two representations which includes also
derivatives of $\Psi(x,y;k)$ up to the second ones if the solutions to SE are considered.

The situations described above are not met however in the two integrable billiards considered but are typical for multisheeted RBRS. Nevertheless to
see how discontinuities influence the respective quantization procedure let us calculate the coefficients of the Fourier expansions for the second
derivatives of $\Psi(x,y;k)$. We have
\be
X_{ij,mn}^{(x^2)}(k)=\frac{1}{ab}\int_{-a}^adx\int_{-b}^bdy\frac{\p^2\Psi(x,y;k)}{\p x^2}f_i\ll(\pi m\frac{x}{a}\r)f_j\ll(\pi n\frac{y}{b}\r)=\nn\\
\frac{4}{ab}\int_0^adx\int_0^bdy\frac{\p^2\Psi^{(ji)}(x,y;k)}{\p x^2}f_i\ll(\pi m\frac{x}{a}\r)f_j\ll(\pi n\frac{y}{b}\r)=\nn\\
\frac{4}{ab}\oint_Ldl\frac{\p\Psi^{(ji)}(x(l),y(l);k)}{\p x}f_i\ll(\pi m\frac{x(l)}{a}\r)f_j\ll(\pi n\frac{y(l)}{b}\r)\cos\alpha_n(l)+\nn\\
(-1)^i\frac{4\pi m}{a^2b}\oint_Ldl\Psi^{(ji)}(x(l),y(l);k)f_{i+1}\ll(\pi m\frac{x(l)}{a}\r)f_j\ll(\pi n\frac{y(l)}{b}\r)\cos\alpha_n(l)-\nn\\
\frac{\pi^2m^2}{a^2}X_{ij,mn}(k)\nn\\
X_{ij,mn}^{(y^2)}(k)=\frac{1}{ab}\int_{-a}^adx\int_{-b}^bdy\frac{\p^2\Psi(x,y;k)}{\p y^2}f_i\ll(\pi m\frac{x}{a}\r)f_j\ll(\pi n\frac{y}{b}\r)=\nn\\
\frac{4}{ab}\oint_Ldl\frac{\p\Psi^{(ji)}(x(l),y(l);k)}{\p y}f_i\ll(\pi m\frac{x(l)}{a}\r)f_j\ll(\pi n\frac{y(l)}{b}\r)\sin\alpha_n(l)+\nn\\
(-1)^j\frac{4\pi n}{ab^2}\oint_Ldl\Psi^{(ji)}(x(l),y(l);k)f_i\ll(\pi m\frac{x(l)}{a}\r)f_{j+1}\ll(\pi n\frac{y(l)}{b}\r)\sin\alpha_n(l)-\nn\\
\frac{\pi^2n^2}{b^2}X_{ij,mn}(k)
\label{12}
\ee
where we have used the Green theorem to express the respective surface integrations by the linear integrals over $L$ - the directed
boundary of the rectangular billiards so that ${\bf n}_p$ denotes the normal to $L$, see Fig.2B.

Taking now however into account that $\Psi_i(x,y;k),\;i=1,...,4$ and therefore $\Psi(x,y;k)$ itself satisfy SE we have further
\be
X_{ij,mn}^{(x^2)}(k)+X_{ij,mn}^{(y^2)}(k)+k^2X_{ij,mn}(k)=0
\label{12a}
\ee
which is equivalent to
\be
\ll(k^2-k^2_{2a2b;mn}\r)X_{ij,mn}(k)=\nn\\
-\frac{4}{ab}\oint_Ldl\ll(\frac{\p\Psi^{(ji)}(x(l),y(l);k)}{\p n}-
\Psi^{(ji)}(x(l),y(l);k)\frac{\p}{\p n}\r)f_i\ll(\pi m\frac{x(l)}{a}\r)f_j\ll(\pi n\frac{y(l)}{b}\r)\equiv\nn\\
x_{ij,mn}(k)\nn\\
k^2_{2a2b;mn}=\frac{\pi^2m^2}{a^2}+\frac{\pi^2n^2}{b^2},\;\;\;\;\; m,n\geq 0
\label{12b}
\ee
The quantities \[k^2_{2a2b;mn}=\frac{\pi^2m^2}{a^2}+\frac{\pi^2n^2}{b^2},\; m,n\geq 0\] will be called further the periodic orbit channel (POC) spectra.

The formulae \mref{12b} give us the direct relations between the coefficients $X_{ij,mn}(k)$ and the boundary conditions which can be put on the solution
$\Psi(x,y;k)$ on the sides of the rectangle. Let us note now that choosing some boundary conditions one fixes the indeces $i,j$ of $\Psi^{(ij)}(x,y)$. Fixing them therefore on $i_0,j_0$ one
decides that $\frac{\p\Psi^{(i_0j_0)}}{\p n}$ or $\Psi^{(i_0j_0)}$ itself vanish on the respective sides of the rectangular billiards. One can then
convince oneself that $x_{j_0i_0,mn}(k)$ in \mref{12} vanish for all $m,n\geq 0$ while  $x_{ji,mn}(k),\;i,j\neq i_0,j_0$ do not necessarily.
Therefore we have in such a case
\be
X_{j_0i_0,mn}(k)(k^2-k^2_{2a2b;mn})=0\nn\\
X_{ji,mn}(k)(k^2-k^2_{2a2b;mn})=x_{ji,mn}(k)\nn\\
i,j\neq i_0,j_0,\;\; m,n\geq 0
\label{13}
\ee

Considering the first of the above equations we are left with the following two possible solutions to it
\begin{enumerate}
\item for any $m,n\geq 0$
\be
k^2\neq k^2_{2a2b;mn},\;\;\;X_{j_0i_0,mn}(k)=0
\label{13a}
\ee
or
\item for some $m_0,n_0$
\be
k^2= k^2_{2a2b;m_0n_0},\;\;\;X_{j_0i_0,mn}(k_{2a2b;m_0n_0})=0,\;(m,n)\neq (m_0,n_0)
\label{14}
\ee
with an arbitrary $X_{j_0i_0,m_0n_0}$.
\end{enumerate}

In the first case we have
\be
\Psi^{(i_0j_0)}(x,y;k)\equiv 0
\label{15}
\ee
while in the second one gets
\be
\Psi^{(i_0j_0)}(x,y;k_{2a2b;m_0n_0})=X_{j_0i_0,m_0n_0}f_{j_0}\ll(\pi m_0\frac{x}{a}\r)f_{i_0}\ll(\pi n_0\frac{y}{b}\r)
\label{16}
\ee
i.e. the standard solution to the energy eigenvalue problem in the rectangle.

For the remaining $\Psi^{(ij)},\;i,j\neq i_0,j_0$ we get in the first case
\be
\Psi^{(ij)}(x,y;k)=\sum_{m,n\geq 0}\frac{x_{ji,mn}(k)}{k^2-k^2_{2a2b;mn}}f_j\ll(\pi m\frac{x}{a}\r)f_i\ll(\pi n\frac{y}{b}\r)\nn\\
0<x<a,\;0<y<b
\label{16a}
\ee
and
\be
\Psi^{(ij)}(x,y;k_{2a2b;m_0n_0})=X_{ji,m_0n_0}f_j\ll(\pi m_0\frac{x}{a}\r)f_i\ll(\pi n_0\frac{y}{b}\r)+\nn\\
\sum_{m,n\neq m_0,n_0}\frac{x_{ji,mn}(k_{2a2b;m_0n_0})}{k^2_{2a2b;m_0n_0}-k^2_{2a2b;mn}}f_j\ll(\pi m\frac{x}{a}\r)f_i\ll(\pi n\frac{y}{b}\r)\nn\\
0<x<a,\;0<y<b
\label{16b}
\ee
in the second one.

Therefore for SPS $\Psi(x,y;k)$ in the rectangular billiards for the chosen boundary conditions $i_0j_0$ we have finally
\be
\Psi^{(FS)}(x,y;k)=\sum_{\ba{l}i,j=1,2\\i,j\neq i_0,j_0\ea}\sum_{m,n\geq 0}\frac{x_{ji,mn}(k)}{k^2-k^2_{2a2b;mn}}f_j\ll(\pi m\frac{x}{a}\r)f_i\ll(\pi n\frac{y}{b}\r)\nn\\
k^2\neq k^2_{2a2b;mn},\;\;\;m,n\geq 0,\;\;-\infty<x<y<+\infty
\label{16c}
\ee
and
\be
\Psi^{(FS)}(x,y;k_{2a2b;m_0n_0})=\sum_{i,j=1,2}X_{ji,m_0n_0}f_j\ll(\pi m_0\frac{x}{a}\r)f_i\ll(\pi n_0\frac{y}{b}\r)+\nn\\
\sum_{\ba{l}i,j=1,2\\i,j\neq i_0,j_0\ea}\sum_{m,n\neq m_0,n_0}\frac{x_{ji,mn}(k_{2a2b;m_0n_0})}{k^2_{2a2b;m_0n_0}-k^2_{2a2b;mn}}f_j\ll(\pi m\frac{x}{a}\r)f_i\ll(\pi n\frac{y}{b}\r)\nn\\
-\infty<x<y<+\infty
\label{16d}
\ee
respectively.

\subsection{SPS and SS in the equilateral triangle billiards}

\hskip+2em The billiards is shown in Fig.2D, its EPP in Fig.2E and its RBRS in Fig.2F. The corresponding SPS $\Psi(x,y;k)$ is defined in EPP by
six solutions $\Psi_i(x,y;k),\;i=1,...,6,$ to SE and continued to RBRS becomes twice periodic with the two
independent periods ${\bf D}_1=[3/2,-\sqrt{3}/2]$ and ${\bf D}_2=[3/2,\sqrt{3}/2]$ so that its Fourier series on RBRS is given by
\be
\Psi^{(FS)}(x,y;k)=\sum_{i,j=1,2}\sum_{m,n\geq 0}\eta_{mn}^{(e)}X_{ij,mn}f_i\ll(2\pi m\frac{x}{3}\r)f_j\ll(2\pi n\frac{y}{\sqrt{3}}\r)\nn\\
\eta_{mn}^{(e)}=\fr(1+(-1)^{m+n})
\label{25a}
\ee
where we have used again the notations from the considerations of the rectangular billiards.

Since $\Psi(x,y;k)$ is periodic also with respect to the periods ${\bf D}_3=[3,0]$ and ${\bf D}_4=[0,\sqrt{3}]$ (see Fig.2F) then
the coefficients in \mref{25a} can be calculated according to Fig.2F using the Green theorem (see App.B) by
\be
\eta_{mn}^{(e)}X_{rs,mn}\ll(k^2-k^2_{3\sqrt{3};mn}\r)=\nn\\
\frac{4}{3\sqrt{3}}\eta_{mn}^{(e)}\ll(k^2-k^2_{3\sqrt{3};mn}\r)\int_0^3dx\int_{-\frac{\sqrt{3}}{2}}^{\frac{\sqrt{3}}{2}}dy\Psi(x,y;k)
f_r\ll(2\pi m\frac{x}{3}\r)f_s\ll(2\pi n\frac{y}{\sqrt{3}}\r)=\nn\\
-\frac{4}{3\sqrt{3}}\eta_{mn}^{(e)}\sum_1^6\oint_{L_i}dl\ll(\frac{\p \Psi_i(x(l),y(l))}{\p n_i}-
\Psi_i(x(l),y(l))\frac{\p }{\p n_i}\r)f_r\ll(2\pi m\frac{x(l)}{3}\r)f_s\ll(2\pi n\frac{y(l)}{\sqrt{3}}\r)=\nn\\
-\frac{4}{3\sqrt{3}}\eta_{mn}^{(e)}\sum_{1\leq i<j\leq 6}\int_{L_{ij}}dl\ll(\frac{\p \Psi_i(x(l),y(l))}{\p n_{ij}}-
\frac{\p \Psi_j(x(l),y(l))}{\p n_{ij}}-\r.\nn\\
\ll.\ll(\Psi_i(x(l),y(l))-\Psi_j(x(l),y(l))\r)\frac{\p}{\p n_{ij}}\r)
f_r\ll(2\pi m\frac{x(l)}{3}\r)f_s\ll(2\pi n\frac{y(l)}{\sqrt{3}}\r)\equiv\nn\\
\eta_{mn}^{(e)}x_{rs,mn}\nn\\
k^2_{3\sqrt{3};mn}=\frac{4\pi^2m^2}{9}+\frac{4\pi^2n^2}{3},\;\;\;\;r,s=1,2
\label{27a}
\ee
where $L_i$ is the directed boundary of the $i$-th triangle, $i=1,...,6,$ while $L_{ij},\;1\leq i<j\leq 6,$ is the directed common side of the
$i$-th and
$j$-th images of EPP including the sides distant by periods of EPP. ${\bf n}_i$ is the normal to $L_i$ and ${\bf n}_{ij}$ - to $L_{ij}$ both
directed outside the $i$-th image according to the Green theorem, see Fig.2F. The functions in the last integrals are the discontinuities of the
gradients of SPS $\Psi(x,y;k)$ normal to the sides and the discontinuities of $\Psi(x,y)$ itself.

The equations \mref{27a} are unique which determine the quantization picture of the equilateral triangular billiards showing explicitly its dependence on
the behaviour of $\Psi(x,y;k)$ on the boundary of the billiards. If however none of such conditions is put on $\Psi(x,y;k)$ and $k^2\neq k^2_{3\sqrt{3};mn}$
for any $m,n\geq 0$ then $\eta_{mn}^{(e)}X_{rs,mn}$ can be calculated by \mref{27a} providing us with SPS of the form
\be
\Psi(x,y;k)=
\sum_{i,j=1,2}\sum_{m,n\geq 0}\eta_{mn}^{(e)}\frac{x_{ij,mn}}{k^2-k^2_{3\sqrt{3};mn}}f_i\ll(2\pi m\frac{x}{3}\r)f_j\ll(2\pi n\frac{y}{\sqrt{3}}\r)\nn\\
k\neq k_{3\sqrt{3};mn},\;\;m,n\geq 0
\label{27b}
\ee
periodic on RBRS and having a continuum of energy spectra.

It is seen also from \mref{27a} that the discrete
values of the energy spectrum for the case are possible only when for some $m,n$ the coefficients $x_{rs,mn},\;r,s=1,2$, in \mref{27a} vanish simultaneously.
For that it is not enough for the normal gradients of $\Psi(x,y;k)$ in \mref{27a} to be continuous on the sides of the billiards separately from the continuity
of $\Psi(x,y;k)$ on the sides and vice versa. Because of that both the discrete spectra, i.e. for the Dirichlet conditions and for the Neumann ones are the
same in the case of the equilateral triangle billiards. This is in accordance with the result got in the semiclassical approach to the problem
\cite{2}

Assuming therefore both the discontinuities of $\Psi(x,y;k)$ vanishing on the triangle sides one can repeat for such a case
the discussion for the rectangular billiards to get for example for $\Psi^{(D)}(x,y;k)$ satisfying on all the sides the Dirichlet boundary conditions
the following results
\begin{itemize}
\item $\Psi(x,y;k)\equiv 0$ if $k^2\neq k^2_{3\sqrt{3};mn}$ for any $m,n\geq 0$; or
\item $k^2=k^2_{3\sqrt{3};m_0n_0}$ for some $m_0,n_0$ and
\be
\Psi(x,y;k_{3\sqrt{3};m_0n_0})=\sum_{i,j=1,2}\eta_{m_0n_0}^{(e)}X_{ij,m_0n_0}f_i\ll(2\pi m_0\frac{x}{3}\r)f_j\ll(2\pi n_0\frac{y}{\sqrt{3}}\r)\nn\\
\Psi^{(D)}(x,y;k_{3\sqrt{3};m_0n_0})=
\eta_{m_0n_0}^{(e)}\ll(X_{21,m_0n_0}\ll(\cos\ll(2\pi m_0\frac{x}{3}\r)\sin\ll(2\pi n_0\frac{y}{\sqrt{3}}\r)+\r.\r.\nn\\
                                                 \cos\ll(\frac{1}{3}\pi m_0(x+\sqrt{3}y)\r)\sin\ll(\frac{1}{\sqrt{3}}\pi n_0(\sqrt{3}x-y)\r)-\nn\\
                                              \ll.\cos\ll(\frac{1}{3}\pi m_0(x-\sqrt{3}y)\r)\sin\ll(\frac{1}{\sqrt{3}}\pi n_0(\sqrt{3}x+y)\r)\r)+\nn\\
                                    X_{11,m_0n_0}\ll(\sin\ll(2\pi m_0\frac{x}{3}\r)\sin\ll(2\pi n_0\frac{y}{\sqrt{3}}\r)-\r.\nn\\
                                                 \sin\ll(\frac{1}{3}\pi m_0(x+\sqrt{3}y)\r)\sin\ll(\frac{1}{\sqrt{3}}\pi n_0(\sqrt{3}x-y)\r)+\nn\\
\ll.\ll.\sin\ll(\frac{1}{3}\pi m_0(x-\sqrt{3}y)\r)\sin\ll(\frac{1}{\sqrt{3}}\pi n_0(\sqrt{3}x+y)\r)\r)\r)
\label{28a}
\ee
\end{itemize}

It follows therefore from \mref{27a} that the Dirichlet boundary conditions is not sufficient to enforce the discreet values
of energy to exist in its spectrum, i.e. non-vanishing discontinuities of the normal gradients of $\Psi(x,y;k)$ on the triangle sides prevents the discrete
values of energy to exist in its spectrum. The same note is valid in the case of the Neumann conditions if $\Psi(x,y;k)$ is discontinuous on the
triangle sides.

Let us note at last that to be freed from the bonds of possible boundary conditions which can be put on $\Psi(x,y;k)$ on the sides of the equilateral
triangle billiards in the considered method one can investigate the respective rhombus-like billiards made by gluing two equilateral triangle ones. Such
a billiards is considered in the next section.

\subsection{SPS and SS in the rhombus-like billiards}

\hskip+2em This case differs essentially from the two previous ones being non-integrable and introducing non-trivial RBRS with branch points absent
in the cases mentioned.
The billiards is shown in Fig.3 with the two sheets of its RBRS in Fig.3D1,2. Both the sheets are periodic with the periods ${\bf D}_x$ and
${\bf D}_y$ while the periods ${\bf D}_i,\;i=1,2,4,5,$ transform one sheet into another. Of course, RBRS for the case as a whole is
invariant on the actions of all the periods mentioned.

The respective SPS $\Psi(x,y)$ is defined on both the sheets by $\Psi_i(x,y),\;i=1,...,6$ - the solutions
to SE in the corresponding images of the rhombus in EPP - by their mirror extensions according to sec.4. $\Psi_i(x,y),\;i=1,...,6$ are defined as usually
by \mref{9}. Denoting by $\Psi^{(1)}(x,y)$ and $\Psi^{(2)}(x,y)$
the branches of $\Psi(x,y)$ on the sheets {\bf 1} and {\bf 2} of Fig.3C respectively we have the following relations between them
\be
\Psi^{(2)}({\bf r})=\Psi^{(1)}({\bf r}-{\bf D}_i),\;\;\;\;i=1,2,4,5
\label{29a}
\ee
and of course also
\be
\Psi^{(i)}({\bf r}+{\bf D}_j)=\Psi^{(i)}({\bf r}),\;\;\;\;i=1,2,\;\;j=x,y
\label{30a}
\ee
where ${\bf r}=(x,y)$.

Since each of the branches is periodic on its sheet they can be represented on them by the following Fourier series
\be
\Psi^{(1,FS)}(x,y)=\sum_{i,j=1,2}\sum_{m,n\geq 0}X_{ij,mn}f_i\ll(2\pi m\frac{x}{3}\r)f_j\ll(2\pi n\frac{y}{\sqrt{3}}\r)=\nn\\
\Psi^{(e,FS)}(x,y)+\Psi^{(o,FS)}(x,y)\nn\\
\Psi^{(e,FS)}(x,y)=\sum_{i,j=1,2}\sum_{m,n\geq 0}\eta_{mn}^{(e)}X_{ij,mn}f_i\ll(2\pi m\frac{x}{3}\r)f_j\ll(2\pi n\frac{y}{\sqrt{3}}\r)\nn\\
\Psi^{(o,FS)}(x,y)=\sum_{i,j=1,2}\sum_{m,n\geq 0}\eta_{mn}^{(o)}X_{ij,mn}f_i\ll(2\pi m\frac{x}{3}\r)f_j\ll(2\pi n\frac{y}{\sqrt{3}}\r)\nn\\
\Psi^{(2,FS)}(x,y)=\Psi^{(1,FS)}(x-\frac{3}{2},y-\frac{\sqrt{3}}{2})=\Psi^{(e,FS)}(x,y)-\Psi^{(o,FS)}(x,y)\nn\\
\eta_{mn}^{(o)}=\fr(1-(-1)^{m+n})
\label{31b}
\ee

It follows from \mref{31b} that $\Psi^{(i)}(x,y),\;i=1,2$, can be defined on a single plane only by
\be
\Psi^{(1)}(x,y)=\Psi^{(e)}(x,y)+\Psi^{(o)}(x,y)\nn\\
\Psi^{(2)}(x,y)=\Psi^{(e)}(x,y)-\Psi^{(o)}(x,y)
\label{31a}
\ee
which the above decompositions are valid in each image of the billiards in RBRS, see Fig.3D3,4.

Thus $\Psi^{(e)}(x,y)$ is periodic on the $x,y$-plane with the periods ${\bf D}_i,\;i=1,2$, see Fig.3D3, while $\Psi^{(o)}(x,y)$ is antyperiodic on
the last periods, i.e. it changes its sign when the periods are applied so that it is periodic with the periods twice as much longer than ${\bf D}_i,\;i=1,2$,
see Fig.3D4. However ${\bf D}_x$ and ${\bf D}_y$ are still the periods for both the functions. Obviously both  $\Psi^{(e)}(x,y)$ and  $\Psi^{(o)}(x,y)$
satisfy SE if  $\Psi^{(i)}(x,y),\;i=1,2,$ do.

Since the decompositions \mref{31a} are valid for each SS $\Psi_i(x,y),\;i=1,...,6$, defined in the respective images on both
the sheets of RBRS there are the following relations between the respective parts of these solutions
\be
\Psi_{i+3}^{(e,o)}(x,y)=\pm\Psi_i^{(e,o)}(x,y)\;\;\;\;i=1,2,3
\label{31b}
\ee
where $"+"$ corresponds to the even parts and $"-"$ to the odd ones.

Moreover we have also in the respective images
\be
\Psi_i^{(e)}(x,y)=\fr(\Psi_i(x,y)+\Psi_{i+3}(x,y)),\;i=1,2,3\;\;\;\;\;\;\;\;\;\;\;\;\;\;\;\;\;\nn\\
\Psi_1^{(o)}(x,y)=\ll\{\ba{lr}
                       \fr(\Psi_1(x,y)-\Psi_4(x,y))&0<x<1,\;0<y\\
                       \fr(\Psi_4(x,y)-\Psi_1(x,y))&0<x<1,\;y<0
                       \ea\r.\nn\\
\Psi_i^{(o)}(x,y)=\fr(\Psi_i(x,y)-\Psi_{i+3}(x,y)),\;\;\;\;i=2,3\;\;\;\;\;\;\;\;\;\;\;\;\;\;\;\;
\label{31c}
\ee

However since $\Psi_i(x,y),\;i=1,4,$ are both smooth passing the cuts joining the sheets also both $\Psi^{(i)}(x,y),\;i=e,o,$ should be
smooth there according to
\be
\Psi^{(o)}(x,0)\equiv\frac{\p\Psi^{(o)}(x,0)}{\p x}\equiv\frac{\p^2\Psi^{(o)}(x,0)}{\p x^2}\equiv 0\nn\\
\ll.\frac{\p\Psi^{(o)}(x,y)}{\p y}\r|_{y=0}\equiv\ll.\frac{\p^2\Psi^{(o)}(x,y)}{\p y^2}\r|_{y=0}\equiv\ll.\frac{\p^2\Psi^{(o)}(x,y)}{\p x\p y}\r|_{y=0}\equiv 0,\nn\\
0<x<1
\label{32a}
\ee
with none condition on $\Psi^{(e)}(x,y)$.

The conditions \mref{32a} determine $\Psi^{(o)}(x,y)$ as the solutions to SE. It is important to consider all of them since  $\Psi^{(o)}(x,y)$ and
their derivatives are discontinuous on the boundaries of the images of the rhombus-like billiards. The respective discontinuities must be therefore
determined by these conditions.

It follows therefore from the above results that for the rhombus-like billiards the respective SPS $\Psi(x,y)$ are of two kinds - the one
$\Psi^{(e)}(x,y)$ and the second $\Psi^{(o)}(x,y)$ which are defined
by \mref{31a}-\mref{31c} and which can be considered completely separately from each other on the single $x,y$-planes shown in Fig.Fig.3D3,4
respectively. In particular it follows from
\mref{31c} that in our further considerations $\Psi_i(x,y),\;i=1,...,6$, can be substituted by $\Psi_i^{(e,o)}(x,y),;i=1,2,3,$
assuming the additional conditions \mref{32a} for $\Psi_1^{(o)}(x,y)$ to be satisfied.

Consider first therefore SPS $\Psi^{(e)}(x,y)$. Its Fourier transform is given by \mref{31b} the coefficients of which are quantized by, see Fig.3D3
\be
\eta_{mn}^{(e)}X_{rs,mn}(k)\ll(k^2-k^2_{3\sqrt{3};mn}\r)=\nn\\
\eta_{mn}^{(e)}\frac{4}{3\sqrt{3}}\ll(k^2-k^2_{3\sqrt{3};mn}\r)\int_0^3dx\int_{-\frac{\sqrt{3}}{2}}^{\frac{\sqrt{3}}{2}}dy
\Psi^{(e)}(x,y;k)f_r\ll(2\pi m\frac{x}{3}\r)f_s\ll(2\pi n\frac{y}{\sqrt{3}}\r)=\nn\\
-\frac{8}{3\sqrt{3}}\eta_{mn}^{(e)}\sum_{p=1}^3\oint_{L_p}dl\ll(\frac{\p \Psi_p^{(e)}(x(l),y(l))}{\p n_p}-
\Psi_p^{(e)}(x(l),y(l))\frac{\p }{\p n_p}\r)\times\nn\\
f_r\ll(2\pi m\frac{x(l)}{3}\r)f_s\ll(2\pi n\frac{y(l)}{\sqrt{3}}\r)\equiv
\eta_{mn}^{(e)}x_{rs,mn}\nn\\
r,s=1,2,\;\;m,n\geq 0
\label{32}
\ee
which the formula is analogous to the one of the equilateral triangle, see \mref{27a}, differing only by the integrations which run by the sides
of the rhombuses rather than of the triangles. Therefore the respective conclusions related to the possible spectra provided by \mref{32} are
analogous to the equilateral triangle billiards, i.e. the spectra of both the cases coincide and are given by the previous section. Therefore the
discrete spectra of $\Psi^{(e)}(x,y)$ are only the POC ones.

Consider next the second SPS $\Psi^{(o)}(x,y)$. Its Fourier transform is again given by \mref{31b} but it must satisfy additionally the conditions
\mref{32a} which in terms of the Fourier coefficients take the following forms
\be
\sum_{k=1,2}{\bf\alpha}_{ik}{\bf X}_{k2}=0\nn\\
\sum_{k=1,2}{\bf\alpha}_{ik}({\bf D}_{k2}^{(x)}-{\bf X}_{k+12}^{(x)})=0\nn\\
\sum_{k=1,2}{\bf\alpha}_{ik}({\bf D}_{k2}^{(y)}-{\bf X}_{k1}^{(y)})=0\nn\\
\sum_{k=1,2}{\bf\alpha}_{ik}({\bf D}_{k2}^{(xy)}-\breve{\bf D}_{k+12}^{(y)}+{\bf X}_{k+11}^{(xy)})=0\nn\\
\sum_{k=1,2}{\bf\alpha}_{ik}({\bf D}_{k2}^{(x^2)}-\check{\bf D}_{k+12}^{(x)}-{\bf X}_{k2}^{(x^2)})=0\nn\\
\sum_{k=1,2}{\bf\alpha}_{ik}({\bf D}_{k2}^{(y^2)}-\check{\bf D}_{k1}^{(y)}-{\bf X}_{k2}^{(y^2)})=0
\label{32b}
\ee
where
\be
({\bf X}_{ij})_{mn}=\eta_{mn}^{(o)}X_{ij,mn}\nn\\
({\bf X}_{ij}^{(x)})_{mn}=(-1)^i\frac{2\pi m}{3}\eta_{mn}^{(o)}X_{ij,mn}\nn\\
({\bf X}_{ij}^{(y)})_{mn}=(-1)^j\frac{2\pi n}{\sqrt{3}}\eta_{mn}^{(o)}X_{ij,mn}\nn\\
({\bf X}_{ij}^{(xy)})_{mn}=(-1)^{i+j}\frac{4\pi^2mn}{3\sqrt{3}}\eta_{mn}^{(o)}X_{ij,mn}\nn\\
({\bf X}_{ij}^{(x^2)})_{mn}=\frac{4\pi^2m^2}{9}({\bf X}_{ij})_{mn}\nn\\
({\bf X}_{ij}^{(y^2)})_{mn}=\frac{4\pi^2n^2}{3}({\bf X}_{ij})_{mn}\nn\\
({\bf D}_{ij}^{(x)})_{mn}=\frac{8}{3\sqrt{3}}\eta_{mn}^{(o)}\sum_{p=1}^3\oint_{L_p}dl\cos\gamma_p\Psi_p^{(o)}(x(l),y(l))
f_i\ll(2\pi m\frac{x(l)}{3}\r)f_j\ll(2\pi n\frac{y(l)}{\sqrt{3}}\r)\nn\\
(\check{\bf D}_{ij}^{(x)})_{mn}=(-1)^i\frac{2\pi m}{3}({\bf D}_{ij}^{(x)})_{mn}\nn\\
({\bf D}_{ij}^{(y)})_{mn}=\frac{8}{3\sqrt{3}}\eta_{mn}^{(o)}\sum_{p=1}^3\oint_{L_p}dl\sin\gamma_p\Psi_p^{(o)}(x(l),y(l))
f_i\ll(2\pi m\frac{x(l)}{3}\r)f_j\ll(2\pi n\frac{y(l)}{\sqrt{3}}\r)\nn\\
(\check{\bf D}_{ij}^{(y)})_{mn}=(-1)^j\frac{2\pi n}{\sqrt{3}}({\bf D}_{ij}^{(y)})_{mn}\nn\\
(\breve{\bf D}_{ij}^{(y)})_{mn}=(-1)^i\frac{2\pi r}{3}({\bf D}_{ij}^{(y)})_{mn}\nn\\
({\bf D}_{ij}^{(xy)})_{mn}=\frac{8}{3\sqrt{3}}\sum_{p=1}^3\oint_{L_p}dl
\cos\gamma_p\frac{\p\Psi_p^{(o)}}{\p y}(x(l),y(l))f_i\ll(2\pi r\frac{x(l)}{3}\r)f_j\ll(2\pi n\frac{y(l)}{\sqrt{3}}\r)\nn\\
({\bf D}_{ij}^{(x^2)})_{mn}=\frac{8}{3\sqrt{3}}\sum_{p=1}^3\oint_{L_p}dl
\cos\gamma_p\frac{\p\Psi_p^{(o)}}{\p x}(x(l),y(l))f_i\ll(2\pi r\frac{x(l)}{3}\r)f_j\ll(2\pi n\frac{y(l)}{\sqrt{3}}\r)\nn\\
({\bf D}_{ij}^{(y^2)})_{mn}=\frac{8}{3\sqrt{3}}\sum_{p=1}^3\oint_{L_p}dl
\sin\gamma_p\frac{\p\Psi_p^{(o)}}{\p y}(x(l),y(l))f_i\ll(2\pi r\frac{x(l)}{3}\r)f_j\ll(2\pi n\frac{y(l)}{\sqrt{3}}\r)\nn\\
({\bf\alpha}_{ij})_{mn}=\int_0^1f_i\ll(2\pi n\frac{x}{3}\r)f_j(2\pi mx)dx,\;\;\;k,j=1,2,\;m,n\geq 0
\label{33}
\ee
where ${\bf n}_p=[\cos\gamma_p,\sin\gamma_p]$ is the normal to $L_p,\;p=1,2,3$. The integrations in \mref{33} run over the sides of the rhombuses
$1,2,3$, of Fig.3D4.

The conditions \mref{32b} should be completed by the respective SE which gives
\be
{\bf D}_{ij}^{(x^2)}+{\bf D}_{ij}^{(y^2)}-\tilde{\bf D}_{i+1j}^{(x)}-\tilde{\bf D}_{ij+1}^{(y)}-{\bf X}_{ij}^{(x^2)}-{\bf X}_{ij}^{(y^2)}+k^2{\bf X}_{ij}=0
\label{34}
\ee

The equations \mref{32b} and \mref{34} provide us with the conditions the coefficients of which defining $\Psi^{(o)}(x,y;k)$ as well as the energy
parameter $k^2$ have to satisfy being quantized. Altogether the equations determine both the energy spectrum and the corresponding SPS
$\Psi^{(o)}(x,y;k)$ for the rhombus billiards.

Let us therefore consider them more closely. The POC spectrum for $\Psi^{(o)}(x,y)$ needs again the latter to be smooth, i.e. all the contour integrals
in \mref{33} must vanish, i.e. ${\bf D}_{ij}^{(k)}=\tilde{\bf D}_{ij}^{(x)}=\tilde{\bf D}_{ij}^{(y)}=0,\;i,j=1,2,\;k=x,y,xy,x^2,y^2$. Therefore
\mref{34} takes the form
\be
(k^2-k^2_{3\sqrt{3};mn})\eta_{mn}^{(o)}X_{ij,mn}=0,\;\;\;\;\;i,j=1,2,\;\;m,n=0,1,2,...
\label{35}
\ee
so that only some $\eta_{m_0n_0}^{(o)}X_{ij,m_0n_0}$ can survive. However
it is easy to check that they also have to vanish if the conditions \mref{32b} are to be satisfied, i.e. $\Psi^{(o)}(x,y)$ vanish totally for their
spectra coinciding with the POC ones.

If however the discontinuities of $\Psi^{(o)}(x,y)$ do not vanish then they can be calculated if $\eta_{mn}^{(o)}X_{ij,mn}$
got from \mref{34} are substituted to the respective conditions \mref{32b}. One gets then also the respective energy
spectrum generated by the vanishing determinant of these conditions.

Therfore we have in the considered case of the billiards the following picture of the energy spectra and
related SPS allowed in the case
\begin{enumerate}
\item the discrete spectra which are divided into three groups
\begin{itemize}
\item the one which coincides with the POC spectra and is identical by both the spectra and the wave functions with the ones for the equilateral
triangle billiards given by \mref{28a} and being insensitive on the chosen boundary conditions - the Dirichlet or the Neumann ones; and
\item the ones which are provided by the solutions to the equations \mref{32b} and \mref{34}, i.e. with the energy spectra
$k_p>0,\;p\geq 1$, and the corresponding SPS $\Psi^{(o)}(x,y;k_p),$ given by
\be
\Psi^{(o,FS)})(x,y;k_p)=\nn\\
\sum_{r,s=1,2}\sum_{m,n\geq 0}\eta_{mn}^{(o)}\frac{x_{rs,mn}(k_p)}{k_p^2-k^2_{3\sqrt{3};mn}}f_r\ll(2\pi m\frac{x}{3}\r)f_s\ll(2\pi n\frac{y}{\sqrt{3}}\r)\nn\\
{\bf x}_{ij}(k_p)=-({\bf D}_{ij}^{(x^2)}(k_p)+{\bf D}_{ij}^{(y^2)}(k_p)-\tilde{\bf D}_{i+1j}^{(x)}(k_p)-\tilde{\bf D}_{ij+1}^{(y)}(k_p))
\label{34b}
\ee
\end{itemize}
\item the continuous energy spectra with their SPS given by \mref{27b};
\end{enumerate}

It is clear that the spectra $k_p>0,\;p\geq 1$, depend on the boundary conditions - the Dirichlet or the Neumann ones which demand vanishing in
the equations \mref{32b}-\mref{34b} the discontinuities of $\Psi^{(o)}(x,y;k_p)$ itself or its normal derivatives respectively.

The lack of the POC spectra in the $k_p>0,\;p\geq 1$ ones of $\Psi^{(o)}(x,y;k_p)$ does not mean that the former do not play any role in the considered
part of the total spectra of the rhombus-like billiards. Namely the POC spectra can still be seen as a source of the resonance behaviour of $\Psi^{(o)}(x,y;k_p)$
when some level of the energy spectra $k_p>0,\;p\geq 1$, is very close to some of the POC one. This resonant behaviour will be discussed in Sec,7.

Finishing this section one has to note that forming by $\Psi^{(e,o)}(x,y;k)$ according to \mref{8} the respective SS to SE satisfying the demanded boundary conditions
on the considered billiards sides one has to use for $\eta$'s the sets of signs shown in the shadowed EPP in Fig.Fig.3D3,4 for the Dirichlet conditions
while for the Neumann ones all the signs must be "$+$" for $\Psi^{(e)}(x,y;k)$ in Fig.3D3 whereas for $\Psi^{(o)}(x,y;k)$ the respective signs are shown
in the shadowed area in Fig.3D4 with the signs closed in the boxes corresponding to the Neumann conditions.

\subsection{SPS and SS in the L-shape billiards}

\hskip+2em For the L-shape
billiards of Fig.Fig.4-5 we have to build their EPP shown in the figures and to define in each of their four images the respective solutions $\Psi_i(x,y),\;i=1,...,4,$.
This can be done using again the forms \mref{9} for each image. Next the solutions can be distributed on the whole
respective RBRS shown in Fig.4D,E and Fig.5D providing us with SPS $\Psi(x,y)$ periodic on RBRS with the periods shown in both the figure. $\Psi(x,y)$
coincides therefore with $\Psi_i(x,y),\;i=1,...,4,$ in the corresponding image of the L-shape billiards in their RBRS.

Further we shall consider only a general case of the L-shape billiards of Fig.5 assuming the relations $a/b$ and $c/d$ to be irrational. It appears that such a case
is simpler than others such as this of Fig.4. The respective RBRS for the case is then built of infinitely many sheets which can be collected into two
families denoted by {\bf a} and {\bf b}
in Fig.5D. Each sheet of the {\bf a}-type is then glued with the {\bf b}-type by a single cut only as it is shown on Fig.5D.
SPS $\Psi(x,y)$ corresponding to the case is of course defined periodically on all the sheets of RBRS. However due to this periodicity on RBRS its
considering can be reduced to two arbitrarily chosen glued sheets of the type {\bf a} and {\bf b} shown on Fig.5D.

Denoting therefore the branch of SPS on the sheet {\bf a} by $\Psi_a(x,y)$ and by  $\Psi_b(x,y)$ the respective branch on the sheet {\bf b} we see that
the first one is periodic on the sheet {\bf a} with the periods ${\bf D}_2=[2b,0]$ and ${\bf D}_3=[0,2c]$ while the second has on the sheet {\bf b} the periods
${\bf D}_1=[2a,0]$ and ${\bf D}_4=[0,2d]$ so that both the branches of  $\Psi(x,y)$ can be represented on their sheets by the respective Fourier series.
Using the previous notations we have
\be
\Psi_a(x,y)=\sum_{i,j=1,2}\sum_{m,n\geq 0}X_{ij,mn}^{(a)}f_i\ll(\pi m\frac{x}{b}\r)f_j\ll(\pi n\frac{y}{c}\r)\nn\\
\Psi_b(x,y)=\sum_{i,j=1,2}\sum_{m,n\geq 0}X_{ij,mn}^{(b)}f_i\ll(\pi m\frac{x}{a}\r)f_j\ll(\pi n\frac{y}{d}\r)
\label{17}
\ee

The series \mref{17} have next to matched. Such a matching however cannot be done on the cut $AB$ which the sheet {\bf a} and {\bf b} of Fig.5D are
glued along since both the series \mref{17} do not converge to $\Psi(x,y)$ there. As we have noticed earlier such a matching can be done by the help
of another Fourier expansion for $\Psi(x,y)$ done in the POC $P_3$ crossing both the sheets {\bf a} and {\bf b}. Since in the latter POC $\Psi(x,y)$
is defined by the periods ${\bf D}_1+{\bf D}_2$ and ${\bf D}_3$ then this expansion is given by
\be
\Psi_{P_3}(x,y)=\sum_{i,j=1,2}\sum_{m,n\geq 0}X_{ij,mn}^{P_3}f_i\ll(\pi m\frac{x}{a+b}\r)f_j\ll(\pi n\frac{y}{c}\r)\nn\\
\label{18}
\ee

Therefore the expansions for $\Psi_{P_3}(x,y)$ and $\Psi_a(x,y)$ are to be matched in the box $R_a$ while  $\Psi_{P_3}(x,y)$ and $\Psi_b(x,y)$ in the
box $R_b$. These matchings must be done up to the second derivatives.

Introducing also the following notation
\be
{\tilde \Psi}_a(x',y)\equiv\Psi_a(x'+b-a,y)\nn\\
{\tilde\Psi}_k(x',y)\equiv\Psi_k(x'+b-a,y)\nn\\
k=1,...,4,\;x'=x-b+a,\;(x,y)\in{\bf a}\nn\\
{\tilde \Psi}_a^{(ij)}(x',y)=
{\tilde\Psi}_1(x',y)+(-1)^j{\tilde\Psi}_2(-x',y)+(-1)^{i+j}{\tilde\Psi}_3(-x',-y)+(-1)^i{\tilde\Psi}_4(x',-y)\nn\\
(x',y)\in R_1\nn\\
\Psi_b^{(ij)}(x,y)=
\Psi_1(x,y)+(-1)^j\Psi_2(-x,y)+(-1)^{i+j}\Psi_3(-x,-y)+(-1)^i\Psi_4(x,-y)\nn\\
(x,y)\in R_2\nn\\
\Psi_{P_3}^{(ij)}(x,y)=
\Psi_1(x,y)+(-1)^j\Psi_2(-x,y)+(-1)^{i+j}\Psi_3(-x,-y)+(-1)^i\Psi_4(x,-y)\nn\\
(x,y)\in R_3
\label{18d}
\ee
we have
\be
{\tilde \Psi}_a(x',y)=\sum_{i,j=1,2}\sum_{m,n\geq 0}{\tilde X}_{ij,mn}^{(a)}f_i\ll(\pi m\frac{x'}{b}\r)f_j\ll(\pi n\frac{y}{c}\r)\nn\\
X_{ij,mn}^{(a)}=(-1)^m\ll({\tilde X}_{ij,mn}^{(a)}\cos\ll(\pi m\frac{a}{b}\r)-(-1)^i{\tilde X}_{i+1j,mn}^{(a)}\sin\ll(\pi m\frac{a}{b}\r)\r)
\label{18e}
\ee

Using the notations similar to the ones applied in the rhombus-like billiards we get for the matching conditions following from \mref{17} and
\mref{18}
\be
{\bf X}_{ij}^{P_3}=\sum_{k=1,2}{\bf\alpha}_{ik}{\bf X}_{kj}^{(a)}+\sum_{k,l=1,2}{\bf\beta}_{ik}{\bf X}_{kC}^{(b)}\gamma_{lj}\nn\\
{\bf D}_{ij}^{P_3,r}-{\bf X}_{i+1j}^{P_3,r}=
\sum_{k=1,2}{\bf\alpha}_{ik}\ll(\tilde{\bf D}_{kj}^{(a,r)}-{\bf X}_{k+1j}^{(a,r)}\r)+
\sum_{k,l=1,2}{\bf\beta}_{ik}\ll({\bf D}_{kC}^{(b,r)}-{\bf X}_{k+1l}^{(b,r)}\r)\gamma_{lj}\nn\\
{\bf D}_{ij}^{P_3,xy}-\breve{\bf D}_{i+1j}^{P_3,y}+{\bf X}_{i+1j+1}^{P_3,xy}=\nn\\
\sum_{k=1,2}{\bf\alpha}_{ik}\ll(\tilde{\bf D}_{kj}^{(a,xy)}-\breve{\tilde{\bf D}}_{k+1j}^{(a,y)}+{\bf X}_{k+1j+1}^{(a,xy)}\r)+
\sum_{k,l=1,2}{\bf\beta}_{ik}\ll({\bf D}_{kC}^{(b,xy)}-\breve{\bf D}_{k+1l}^{(b,y)}+{\bf X}_{k+1l+1}^{(b,xy)}\r)\gamma_{lj}\nn\\
{\bf D}_{ij}^{P_3,r^2}-\check{\bf D}_{i+1j}^{P_3,r}-{\bf X}_{ij}^{P_3,r^2}=\nn\\
\sum_{k=1,2}{\bf\alpha}_{ik}\ll(\tilde{\bf D}_{kj}^{(a,r^2)}-\check{\tilde{\bf D}}_{k+1j}^{(a,r)}-{\bf X}_{kj}^{(a,r^2)}\r)+
\sum_{k,l=1,2}{\bf\beta}_{ik}\ll({\bf D}_{kC}^{(b,r^2)}-\check{\bf D}_{k+1l}^{(b,r)}-{\bf X}_{kC}^{(b,r^2)}\r)\gamma_{lj}\nn\\
r=x,y
\label{18a}
\ee
where the respective quantities entering the above equations are defined in App.C.

One needs also to add to the above equations the ones which follow from SE satisfied by the branches \mref{17} and \mref{18}. They are
\be
\tilde{\bf D}_{ij}^{(a,x^2)}-\check{\tilde{\bf D}}_{i+1j}^{(a,x)}+\tilde{\bf D}_{ij}^{(a,y^2)}-\check{\tilde{\bf D}}_{i+1j}^{(a,y)}+
k^2{\bf X}_{ij}^{(a)}-{\bf X}_{ij}^{(a,x^2)}-{\bf X}_{ij}^{(a,y^2)}=0\nn\\
{\bf D}_{ij}^{(b,x^2)}-\check{\bf D}_{i+1j}^{(b,x)}+{\bf D}_{ij}^{(b,y^2)}-\check{\bf D}_{i+1j}^{(a,y)}+
k^2{\bf X}_{ij}^{(b)}-{\bf X}_{ij}^{(b,x^2)}-{\bf X}_{ij}^{(b,y^2)}=0\nn\\
{\bf D}_{ij}^{P_3,x^2}-\check{\bf D}_{i+1j}^{P_3,x}+{\bf D}_{ij}^{P_3,y^2}-\check{\bf D}_{i+1j}^{P_3,y}+
k^2{\bf X}_{ij}^{P_3}-{\bf X}_{ij}^{P_3,x^2}-{\bf X}_{ij}^{P_3,y^2}=0\nn\\
\label{18c}
\ee

It is seen that \mref{18a} and \mref{18c} form a system of the homogeneous equations for the {\bf D}- and {\bf X}-type quantities which must be
determined by it together with the energy spectra which are determined by the vanishing determinant of the system. The system takes different forms
depending on the boundary conditions demanded for $\Psi(x,y;k)$ - in the case of the Dirichlet conditions the quantities
${\bf D}_{ij}^{(k,l)},\;k=a,b,\;{\bf D}_{ij}^{P_3,l},\;\;l=x,y$, must vanish in the equations of the system while for the Neumann conditions
these are the quantities ${\bf D}_{ij}^{(k,l)},\;k=a,b,\;{\bf D}_{ij}^{P_3,l},\;\;l=x^2,y^2,xy$, which are demanded to vanish instead. Therefore
this difference must produce also different energy spectra for different conditions demanded.

As in the case of the rectangular billiards the possible boundary conditions which can be put on SS in the L-shape billiards being compatible with EPP
of Fig.5B corresponds to the following four sets of the signs $\eta_i,\;i=1,...,4$, in \mref{8}: $(++++),(+--+),(++--),(+-+-)$. The latter correspond to
the following Neumann (N) or Dirichlet (D) boundary conditions for the horizontal and the vertical sides of the L-shape billiards respectively:
(NN)$\equiv(22)$, (ND)$\equiv(21)$, (DN)$\equiv(12)$, (DD)$\equiv(11)$ by which we are adopting our earlier convention for the rectangular billiards.

An analysis of the relations \mref{18a} and \mref{18c} to see whether they can produce the superscar states will be done in Sec.7.

Let us note by the way that the case of the billiards considered by Richens and Berry \cite{8} are just a particular reduction to its half of the
L-shape billiards symmetric with respect to its diagonal $AB$, see Fig.5A and therefore can be treated by the method applied in our paper.

\subsection{SPS and SS in the Bogomolny-Schmit billiards}

\hskip+2em This case deserves for considerations due to its fame \cite{6}. However its treatment goes along the similar ways as those in
the previous cases particularly this developed in the case of the L-shape billiards. Fig.6 shows clearly why the case is similar to the L-shape one. Namely there
are also two basic sheets {\bf a} and {\bf b} on which the respective SPS $\Psi(x,y)$ is defined being repeated by respective periods on the
remaining sheets infinite in a number. Following the procedure used in the L-shape billiards one gets for the branches $\Psi_a(x,y)$ and $\Psi_b(x,y)$ of $\Psi(x,y)$ on the respective sheets
\be
\Psi_a(x,y)=\sum_{i,j=1,2}\sum_{m,n\geq 0}X_{ij,mn}^{(a)}f_i\ll(\pi m\frac{x}{b}\r)f_j\ll(\pi n\frac{y}{a}\r)\nn\\
\Psi_b(x,y)=\sum_{i,j=1,2}\sum_{m,n\geq 0}\eta_{mn}^{(e)})X_{ij,mn}^{(b)}f_i\ll(\pi m\frac{x}{\sqrt{2}b}\r)f_j\ll(\pi n\frac{y}{\sqrt{2}a}\r)
\label{19}
\ee
The above two branches of $\Psi(x,y)$ can be then matched by helping with the POC {\bf P} shown in Fig.6Ea,b as the shadowed area. The POC is
periodic along the $x$-axis with the periods ${\bf D}_P=[2a,0]$, and has the width $2b$ so that the Fourier series of $\Psi(x,y)$ in the POC is the
following
\be
\Psi_P(x,y)=\sum_{i,j=1,2}\sum_{m,n\geq 0}X_{ij,mn}^Pf_i\ll(\pi m\frac{x}{a}\r)f_j\ll(\pi n\frac{y}{b}\r)\nn\\
\label{21}
\ee

The respective procedure of matching $\Psi_a(x,y)$ and $\Psi_b(x,y)$ by $\Psi_P(x,y)$ is very similar to the one applied in the case of the L-shaped
billiards so that we do not repeat it fully hear. Instead we only
invoke the respective results demanded by SE which must be satisfied by the branches $\Psi_a(x,y)$ and $\Psi_b(x,y)$ of $\Psi(x,y)$ which are the
following
\be
X_{rs,mn}^{(a)}\ll(k^2-k^2_{2b2a;mn}\r)=\nn\\
-\frac{1}{ab}\sum_{i=1}^{16}\oint_{L_i^{(a)}}dl\ll(\frac{\p \Psi_i(x(l),y(l))}{\p n_i}-\Psi_i(x(l),y(l))\frac{\p}{\p n_i}\r)
f_r\ll(\pi m\frac{x(l)}{b}\r)f_s\ll(\pi n\frac{y(l)}{a}\r)\equiv\nn\\
x_{rs,mn}^{(a)}\nn\\
\eta_{mn}^{(e)}X_{rs,mn}^{(b)}\ll(k^2-k^2_{2\sqrt{2}b2\sqrt{2}a;mn}\r)=\nn\\
\eta_{mn}^{(e)}\sum_{\ba{c}i=1\\i\neq 4,5,12,13\ea}^{16}\oint_{L_i^{(b)}}dl\ll(\frac{\p \Psi_i(x(l),y(l))}{\p n_i}-
\Psi_i(x(l),y(l))\frac{\p}{\p n_i}\r)\times\nn\\
f_r\ll(\pi m\frac{x(l)}{\sqrt{2}b}\r)f_s\ll(\pi n\frac{y(l)}{\sqrt{2}a}\r)\equiv \eta_{mn}^{(e)}x_{rs,mn}^{(b)}\nn\\
X_{rs,mn}^P\ll(k^2-k^2_{2a2b;mn}\r)=\nn\\
-\frac{1}{ab}\sum_{i=1}^{16}\oint_{L_i^P}dl\ll(\frac{\p \Psi_i(x(l),y(l))}{\p n_i}-\Psi_i(x(l),y(l))\frac{\p}{\p n_i}\r)
f_r\ll(\pi m\frac{x(l)}{a}\r)f_s\ll(\pi n\frac{y(l)}{b}\r)\equiv\nn\\
x_{rs,mn}^P\nn\\
r,s=1,2
\label{22}
\ee
where all the integrations run inside the respective boxes with gratings for $L_i^{(a)},L_i^{(b)}$ and the shadowed one for $L_i^P$ shown in Fig.6Eab.

The relations \mref{22} are the basic ones for the discussion in Sec.7 of the role of POCs in the superscar properties of $\Psi(x,y)$.

As in the case of the L-shaped billiards if $k^2\neq k^2_{2i2j;mn},\;i,j=b,a;\;\sqrt{2}b,\sqrt{2}a;\;a,b$, for any $m,n$ then the equations
\mref{22} together with the ones got by matching $\Psi_a(x,y)$ and $\Psi_b(x,y)$ by $\Psi_P(x,y)$ becomes again the infinite linear system of
homogeneous equations for the coefficients $X_{ij,mn}^{(k)},\;X_{ij,mn}^P,\;k=a,b,$ and the discontinuities of $\Psi(x,y)$ the nonzero solutions for
which demand vanishing of the respective determinant of the system. The latter
condition defines then the discrete spectrum of energy for the considered billiards. The cases when $k^2= k^2_{2i2j;mn}$ for some $m,n$, i.e. the existence
of the POC spectra in the Bogomolny-Schmit billiards will be considered in sec.7.

One can easily check that possible boundary conditions which can be put on the sides $a,b,c$ of the triangle allowed by EPP corresponding to the
case are the following $(N,N,N)$, $(D,D,D)$, $(D,D,N)$ and $(N,N,D)$ where $N$ stands for the Neumann condition while $D$ - for the Dirichlet one which
correspond to the following sets of the signs $\eta_i,\;i=1,...,16$ in \mref{8}: $++++...$, $+-+-...$, $++--...$ and $+--+...$ where these first
four signs are repeated in each of the sequences in the place of dots.

\subsection{SPS and SS in the rectangular billiards with a rectangular hole}

\hskip+2em  This is the last example considered in the paper, see Fig.7. It differs from the previous ones by the number of sheets of RBRS necessary to build on it
the respective SPS and SS and equal to six. The direct consequences of the increased number of sheets of RBRS is a number of branches of SPS equal
also to six. However the basic steps in the constructions of SPS and SS as well as
of the relations between respective branches remains the same. Namely according to Fig.7E we can write the following Fourier representations for the
respective branches

\be
\Psi^{(A)}(x,y)=\sum_{i,j=1,2}\sum_{m,n\geq 0}X_{ij,mn}^Af_i\ll(\pi m\frac{x}{w}\r)f_j\ll(\pi n\frac{y}{b}\r)\nn\\
\Psi^{(B)}(x,y)=\sum_{i,j=1,2}\sum_{m,n\geq 0}X_{ij,mn}^Bf_i\ll(\pi m\frac{x}{f}\r)f_j\ll(\pi n\frac{y}{b}\r)\nn\\
\Psi^{(k)}(x,y)=\sum_{i,j=1,2}\sum_{m,n\geq 0}X_{ij,mn}^{(k)}f_i\ll(\pi m\frac{x}{c}\r)f_j\ll(\pi n\frac{y}{h}\r)\nn\\
\Psi^{(l)}(x,y)=\sum_{i,j=1,2}\sum_{m,n\geq 0}X_{ij,mn}^{(l)}f_i\ll(\pi m\frac{x}{c}\r)f_j\ll(\pi n\frac{y}{e}\r)\nn\\
k=1,2,\;l=3,4
\ee
which can be matched by the respective POCs $P_3$ and $P_4$ on which the Fourier series of $\Psi^{(A)}(x,y)$ are defined by
\be
\Psi^{P_3}(x,y)=\sum_{i,j=1,2}\sum_{m,n\geq 0}X_{ij,mn}^{P_3}f_i\ll(\pi m\frac{x}{a}\r)f_j\ll(\pi n\frac{y}{h}\r)\nn\\
\Psi^{P_4}(x,y)=\sum_{i,j=1,2}\sum_{m,n\geq 0}X_{ij,mn}^{P_4}f_i\ll(\pi m\frac{x}{a}\r)f_j\ll(\pi n\frac{y}{e}\r)
\label{25}
\ee

The matching conditions provided by the representations \mref{24}-\mref{25} are constructed along the same line as in the case of the L-shape
billiards producing $2\times 2\times 5$ linear homogeneous matrix equations plus the six one which follow from SE. We quote only the latter necessary
for the later discussion. They are
\be
X^{(J)}_{rs,mn}(k^2-k_{2p_J2q_J,mn}^2)=\nn\\
\frac{1}{p_Jq_J}\oint_{L_J}dl\ll(\frac{\p\Psi_J^{(Y_sY_r)}(x,y)}{\p n}-
\Psi_J^{(Y_sY_r)}(x,y)\frac{\p }{\p n}\r)f_r\ll(\pi m\frac{x}{b}\r)f_s\ll(\pi n\frac{y}{c}\r)\equiv x_{rs,mn}^{(J)}\nn\\
J=A,B,\;\;r,s=1,2,\;\;m,n\geq 0,\;\;p_A=w,\;p_B=f,\;q_A=q_B=b\nn\\
{\tilde X}^{(i)}_{rs,mn}(k^2-k_{2p_i2q_i,mn}^2)=\nn\\
\frac{1}{p_iq_i}\oint_{L_i}dl\ll(\frac{\p{\tilde\Psi}_i^{(Y_sY_r)}(x',y)}{\p n}-
{\tilde\Psi}_i^{(Y_sY_r)}(x',y)\frac{\p }{\p n}\r)f_r\ll(\pi m\frac{x}{b}\r)f_s\ll(\pi n\frac{y}{c}\r)\equiv x_{rs,mn}^{(i)}\nn\\
i=1,2,3,4\;\;r,s=1,2,\;\;m,n\geq 0,\;\;p_1=...=p_4=c,\;q_1=q_2=h,\;q_3=q_4=e\nn\\
X^{P_i}_{rs,mn}(k^2-k_{2a2q_i,mn}^2)=\nn\\
\frac{1}{aq_i}\oint_{L_{P_i}}dl\ll(\frac{\p\Psi_i^{(Y_sY_r)}(x,y)}{\p n}-
\Psi_i^{(Y_sY_r)}(x,y)\frac{\p }{\p n}\r)f_r\ll(\pi m\frac{x}{a}\r)f_s\ll(\pi n\frac{y}{q_i}\r)\equiv x_{rs,mn}^{P_i}\nn\\
i=3,4,\;\;r,s=1,2,\;\;m,n\geq 0,\;\;q_3=h,\;q_4=e\nn\\
\label{26}
\ee
where the following notations are used
\be
{\tilde \Psi}^{(l)}(x',y)\equiv\Psi^{(l)}(x'+(-1)^lw,y)\nn\\
{\tilde\Psi}_k^{(l)}(x',y)\equiv\Psi_k(x'+(-1)^lw,y)\nn\\
{\tilde \Psi}_l^{(Y_iY_j)}(x',y)=\nn\\
{\tilde\Psi}_1^{(l)}(x',y)+(-1)^{j+1}{\tilde\Psi}_2^{(l)}(-x',y)+(-1)^{i+j}{\tilde\Psi}_3^{(l)}(x',-y)+(-1)^{i+1}{\tilde\Psi}_4^{(l)}(-x',-y)\nn\\
(x',y)\in {\bf l},\;l,k=1,...,4\nn\\
X_{rs,mn}^{(l)}={\tilde X}_{rs,mn}^{(l)}\cos\ll(\pi m\frac{w}{c}\r)+(-1)^{r+l}{\tilde X}_{r+1s,mn}^{(l)}\sin\ll(\pi m\frac{w}{c}\r)\nn\\
l=1,2,3,4\;r,s=1,2\nn\\
\Psi_J^{(Y_iY_j)}(x,y)=
\Psi_1(x,y)+(-1)^{j+1}\Psi_2(-x,y)+(-1)^{i+j}\Psi_3(x,-y)+(-1)^{i+1}\Psi_4(-x,-y)\nn\\
i,j=1,2,\;(x,y)\in{\bf J},\;J=A,B,\;Y_1=N,\;Y_2=D
\label{27}
\ee

The procedure of calculating of the non-POC spectra are the same as previously - the equations \mref{26} together with the 20 matrix ones not
written but
containing the respective discontinuities of $\Psi(x,y)$ form the system of the linear homogeneous equations the determinant of which when vanishing
determines the energy spectra and $\Psi(x,y)$ itself for the case. The existence of the respective POC spectra will be discussed in the
next section.

\section{Superscar states (SSS) in POCDRB}

\hskip+2em In the previous sections we have shown clearly the role played by POCs in the constructions of RBRS and of SPS as well as in the quantizations
of the latter in POCDRB. It was shown in particular that SPS and their corresponding energy spectra in the billiards considered are determined by the equations which
contain the contributions from the POC spectra as their essential ingredients. The latter spectra fully construct the ones in the integrable cases of the
rectangular and equilateral triangle billiards while in the remaining cases they may constitute an exceptional separate part of their discrete spectra. The
question arises whether it is also possible according to the equations they have to satisfy and what kind of states they generate. Bogomolny and Schmit
suggested \cite{6}-\cite{7} that POCs themselves can generate states called superscars states (SSS) which superpose with the other in the rational billiards. However
our results show much more, namely that SSS independently of whether they exist or not cannot be a problem which can be considered separately from
the main one, i.e. from the quantization of the
rational billiards being its immanent part. To justify further this statement we shall discuss the existence of the solutions of the quantization equations
corresponding
to the different rational billiards considered in the paper and corresponding to the POC spectra of the billiards cases.

\subsection{POCs and SSS in the integrable rational billiards}

\hskip+2em According to our definition of POCs in the present paper in the cases now considered their RBRS are simply planes, i.e. each RBRS is deprived
of any branching point and
also of any singular diagonal. Therefore each bundle of trajectories parallel to some period of the case considered constitute a POC, i.e. each POC in the
considered cases covers totally the respective plane having no boundaries. However SPS constructed on such a POC must be periodic not only by the period of the POC
but also by other periods not parallel to the POC. But it is not always possible to find a second independent period which together with the POC one
could reproduce the full space of periods of RBRS, i.e. in such cases of POCs a set of SSS constructed on them would coincide only with subsets of the sets
\mref{10}
and \mref{25a} of the respective SPS. Therefore there is also a limited set of POCs which SSS constructed on can reproduce all SPS related with both the
cases. In the case of the rectangular billiards and of the equilateral triangle one to such a set belong the POCs $P_x,\;P_y$ with their trajectories
parallel to the $x$- or $y$-axes respectively.

\subsection{SSS in the rhombus-like billiards}

\hskip+2em In sec.6.3 it was noticed that the discrete energy spectra determined by \mref{32b} and \mref{34} are divided into the two classes - the
one of the states $\Psi^{(e)}(x,y)$ the spectra of which coincide with the POC ones and corresponded to the spectra of the POCs $P$ and $P'$ and the
other of the states $\Psi^{(e)}(x,y)$ which cannot contain spectra of any POC one. However some properties of SSS can be still visible also in the later case
of the states $\Psi^{(o)}(x,y)$ as a kind of resonant effects provided by the quantization conditions \mref{32b} and \mref{34} and leading to SPS
given by \mref{34b}. Namely, since
the eigenvalues $k_p,\;p\geq 1$, are of course discrete and distributed along the positive $k$-axis as well as $k_{3\sqrt{3};mn},\;m,n\geq 0$ then
if $k_p$ is close to some $k_{3\sqrt{3};m_0n_0}$ for some $p_0$ the term with the coefficient $\frac{x_{rs,mn}(k_{p_0})}{k_{p_0}^2-k^2_{3\sqrt{3};m_0n_0}}$ is
large and can dominate over the rest of the series in \mref{34b} in the whole domain of EPP in Fig.3D4. The dominating term has exactly
the form of SSS $\Psi^P_{m_0n_0}(x,y)$ defined on the POC $P$, see Fig.3D1.

We can conclude therefore that in the considered billiards there are "pure" SSS as well as the states which can imitate SSS if their
eigenvalues are sufficiently close to the POC ones.

\subsection{SSS in the L-shape billiards}

\hskip+2em Assuming the Dirichlet boundary conditions and the smooth behaviour of $\Psi(x,y)$ on RBRS we get from \mref{18c} and \mref{C1}
\be
{\tilde X}_{22,mn}^{(a)}\ll(k^2-k^2_{2b2c;mn}\r)=
(-1)^m\frac{\pi m}{b^2c}\int_0^cdy{\tilde\Psi}_a^{(DD)}(-b,y)\sin\ll(\pi n\frac{y}{c}\r)\nn\\
X_{22,mn}^{(b)}\ll(k^2-k^2_{2a2d;mn}\r)=-(-1)^m\frac{\pi m}{a^2d}\int_0^cdy\Psi_b^{(DD)}(a,y)\sin\ll(\pi n\frac{y}{d}\r)\nn\\
{\tilde\Psi}_a^{(DD)}(-b,y)=\Psi_b^{(DD)}(a,y),\;\;\;0<y<c\nn\\
X_{22,mn}^{P_3}\ll(k^2-k^2_{2e2c;mn}\r)=-(-1)^n\frac{\pi m}{c^2e}\int_0^adx\Psi_{P_3}^{(DD)}(x,c)\sin\ll(\pi m\frac{x}{e}\r)
\label{39}
\ee

If $k^2$ is to coincide with some value of the POC spectra present in the first two equations in \mref{39} then it has to be
$\Psi_a^{(DD)}(-a,y)=\Psi_b^{(DD)}(a,y)=0,\;0<y<c$, i.e. on the vertical $l_v$ segment of the L-shape billiards of Fig.5A respectively.
Then $k^2=k^2_{2b2c;m_0n_0}$ or $k^2=k^2_{2a2d;m_0n_0}$ for some $m_0,n_0$ and one of the case excludes the other because of
the irrationality of the sides. If $k^2=k^2_{2b2c;m_0n_0}$ then ${\tilde X}_{22,m_0n_0}^{(a)}\neq 0$ while ${\tilde X}_{22,mn}^{(a)}=0$
for $m,n\neq m_0,n_0$ and $X_{22,mn}^{(b)}=0$ for all $m,n\geq 0$. Therefore we have
\be
\Psi^{(DD)}_{a;m_0n_0}(x,y)=(-1)^m{\tilde X}_{22,m_0n_0}^{(a)}\sin\ll(\pi m_0\frac{x+a}{b}\r)\sin\ll(\pi n_0\frac{y}{c}\r)\nn\\
-a<x<b-a,\;0<y<c\nn\\
\Psi^{(DD)}_{b;m_0n_0}(x,y)=0,\;\;\;\;-2a<x<-a,\;0<y<d
\label{40}
\ee

The remaining coefficients $X_{ij,mn}^{(l)}(k_{2b2c;m_0n_0}),\;i,j=1,2,\;l=a,b$, defining $\Psi(x,y;k_{2b2c;m_0n_0})$ for the case can be calculated
by the equations \mref{18a} being now inhomogeneous and providing them to be proportional to ${\tilde X}_{22,m_0n_0}^{(a)}$ each.

Therefore in the L-shape billiards such a mode seems to develop SSS $\Psi^{P_1}_{m_0n_0}(x,y)$ shown in Fig.5E2. Unfortunately the respective SS
$\Psi^{(DD)}(x,y;k_{2b2c;m_0n_0})$ defined by
\be
\Psi^{(DD)}(x,y;k_{2b2c;m_0n_0})=\ll\{\ba{lr}
                                      \Psi^{(DD)}_{a;m_0n_0}(x,y)&-a<x<b-a,\;0<y<c\\
                                      \Psi^{(DD)}_{b;m_0n_0}(x,y)=0&-2a<x<-a,\;0<y<d
                                      \ea\r.
\label{40a}
\ee
cannot exist in the L-shape billiards since its derivative with respect to $x$ at $x=-a,\;0<y<c$, i.e. inside the billiards is discontinuous.

Analyses of other modes of the POC spectra which follow from \mref{39} lead us mutatis mutandis to the similar conclusions that also other SSS
such as shown in Fig.Fig.5E1-5E4 cannot exist in the L-shape billiards for the similar reasons.

Nevertheless the resonance effect of POC spectra discussed in the case of the rhombus-like billiards can be observed also in the L-shape one by
the same reasons and are shown schematically in Fig.Fig.5E5-5E8

However the above conclusions are not true if the L-shape billiards is DRPB, i.e. if their horizontal and vertical periods are commensurate in
each of their groups as in the case of the L-shape billiards of Fig.4. Then some energy levels of the spectra corresponding to the POCs of
Fig.Fig.5E1-5E4 can be tuned with each other providing us with the superscars solutions \cite{6}-\cite{7},\cite{10} which are shown schematically in Fig.5E9.

\subsection{SSS in the Bogomolny-Schmit billiards}

\hskip+2em  POCs with their SSS which can be distinguished in the considered case are ${\bf P}_a,{\bf P}_b,{\bf P}_{b'}$. ${\bf P}_a$ constitute the
sheet {\bf a} while ${\bf P}_b$ and ${\bf P}_{b'}$ - the sheet {\bf b}. $\Psi_a(x,y)$ is defined on the respective POC ${\bf P}_a$ directly while
$\Psi_b(x,y)$ is defined on both the POCs
${\bf P}_b$ and ${\bf P}_{b'}$. Therefore both SS can become SSS if they are found as solutions for the POC spectra present in \mref{22}. Let us
check such a possibility.

First the rho in \mref{22} have to vanish. This can be achieved looking for SS which are smooth on the side traces of the triangle in EPP but also
demanding the respective smoothness of these SS
on the cuts $AA'$ and $BB'$ in Fig.6Ea as the necessary conditions for the coefficients $x_{rs,mn}^{(i)},\;i=a,b$, in \mref{22} to vanish. If all
these take place then
\begin{enumerate}
\item $\Psi_a(x,y)$ and $\Psi_b(x,y)$ vanish identically on their sheets for $k^2\neq k^2_{2b2a;mn},k^2_{2\sqrt{2}b2\sqrt{2}a;mn}$, $\;m,n\geq 0$;
\item $\ll\{\ba{lr}
        k^2=k^2_{2b2a;m_0n_0}&\\
     \Psi_a(x,y)=\sum_{i,j=1,2}X_{ij,m_0n_0}^{(a)}f_i\ll(\pi m_0\frac{x}{b}\r)f_j\ll(\pi n_0\frac{y}{a}\r)&\\
       \Psi_b(x,y)\equiv 0&
            \ea\r.$
\item $\ll\{\ba{lr}
        k^2=k^2_{2\sqrt{2}b2\sqrt{2}a;m_0n_0}&\\
       \Psi_a(x,y)\equiv 0&\\
       \Psi_b(x,y)=\fr(1+(-1)^{m_0+n_0})
       \sum_{i,j=1,2}X_{ij,m_0n_0}^{(b)}f_i\ll(\pi m_0\frac{x}{\sqrt{2}b}\r)f_j\ll(\pi n_0\frac{y}{\sqrt{2}a}\r)&
           \ea\r.$
\end{enumerate}

It is however easy to note that both the above solutions are not acceptable for the same reasons as in the previous case. Such a conclusion seems
to be in disagreement with the results of Bogomolny $et\; al$ who claim that they observe the effect 2. above of SSS in their triangle billiards
\cite{6}. Since it was not the case then what their calculations did present really?

The answer is similar to the one got in the L-shape billiards case, i.e. it was a kind of the resonant effect when some energy $k$ of the spectra
given by \mref{21}-\mref{22} was close (in fact very close \cite{3}-\cite{6}) to some $k$ of the POC spectra. The respective analyses of such a case is similar to the one of the L-shape
billiards so that we do not repeat it here.

\subsection{SSS in the rectangular billiards with the parallel hole}

\hskip+2em The respective analysis corresponding to the case strongly reminds the one for the L-shape billiards with the analogous conclusions.
Namely, SSS corresponding to the POCs shown in Fig.7C cannot exist parallel to other states of the considered billiards if their dimensions are
not commensurate in which the cases POCs can be observed as resonant effects exclusively. Only when the horizontal dimensions and the vertical ones
of the billiards considered are commensurate in their groups then there are possibilities of tunning the modes of SSS corresponding to the POCs
shown in Fig.7C to get the global effect of SSS in the billiards.

\subsection{SSS in the rectangular billiards with the rotated hole}

\hskip+2em In this case the role of POCs in the quantization of the billiards is even more spectacular then in the previous ones. A multitude of
POCs present in the case generate the respective multitude of sheets of the corresponding RBRS, see Fig.7C(A-H). This provide us with the corresponding
number of the POC spectra entering the quantization conditions defining the case. While it is rather unreal to expect "pure" superscar states as
SS in the considered case their exposing as resonances in the possible eigenvalue states should be abundant.

\section{Summary and discussion}

\hskip+2em In our paper utilizing the rational form of RB expressed by the possibility of construction of the respective EPP and their periodic
continuations to RBRS we have formulated the procedure of quantization of these billiards allowing us for drawing a number of conclusions about
the properties of the wave functions and energy spectra of the considered billiards some of them being general.

First we have shown in Sec.Sec.2-4 that
\begin{itemize}
\item any stationary solution (SS) to SE in RB can be continued on its whole EPP and next - by the periodic continuation - on the whole RBRS
showing in this way the existence of periodic stationary solutions to SE on RBRS with the definite energy;
\item the above procedure of continuation of SS into RBRS can be inverted by constructing the stationary pre-solutions (SPS) to SE on RBRS by which
SS is next obtained by simple algebra;
\end{itemize}

In the same sections it was shown that EPP itself is enough already for the most general formulation of the quantization conditions for SPS in RB
and its energy spectra proving the following main conclusions of the paper
\begin{itemize}
\item both the energy spectra and the respective SPS which are provided by the quantization on EPP are
completely determined by the periods of EPP;
\item the last conclusion is valid also for SS in RB built by SPS.
\end{itemize}

The fruitful way of getting more specific results for the considered subject of the RB quantization was distinguishing the class of them EPP of which
could be decomposed by POCs (POCDRB). By this the role played by these classical objects in the RB quantization noticed by Bogomolny and Schmit \cite{6}
was fully taken into account. It then appeared that the respective RBRS can be constructed in the standard way by plane sheets glued between
themselves along cuts - the constructions well known from the complex
analysis. RBRS made in this way appeared to be finitely sheeted for doubly rational PB (DRPB)
and infinitely sheeted for other POCDRB. On such RBRS the following steps were possible
\begin{enumerate}
\item constructions of SPS by expanding them into Fourier series on each sheet - branches of SPS on different sheets could then have different pairs
of periods solving a potential problem of having by SPS more than two independent periods on a plane;
\item writing the respective quantization conditions as the ones which determine not only the respective energy spectra but also discontinuities of
SPS on the image sides;
\item explicitely including into the quantization procedure spectra of POCs which sheets were glued of;
\item a possibility of discussing of an existence of states with the POC spectra known as the superscars states (SSS).
\end{enumerate}

Just the last possibility allowed us to conclude in Sec.7 that
\begin{enumerate}
\setcounter{enumi}{4}
\item the superscar states of Bogomolny and Schmit \cite{6} can exist as the "pure"
states only in a very limited number of RB - mostly they can manifest themselves as resonances if some cases of their energies are close to the ones
of RB considered.
\end{enumerate}

The conclusion similar to the last one was also drawn in our earlier paper \cite{3}.

As we have noticed in Introduction the question whether every EPP can be decomposed into some system of parallel POCS seems to be open and even it is
not known what a subset of all PB the set of POCDRB is. Nevertheless one can suppose that some of the quantum properties of POCDRB established in
the paper and set in the points 1.-5. above are common for all RB.

\appendix

\section{Forming solutions to the Schr\"odinger equation in RB}

\hskip+2em In any image of EPP a general solution to the equation \mref{4} can be given in the following form
\be
\Psi(x,y;k)=\int_0^{2\pi}(C(\phi)\cos(kx\cos\phi)\cos(ky\sin\phi)+D(\phi)\cos(kx\cos\phi)\sin(ky\sin\phi)+\nn\\
                              E(\phi)\sin(kx\cos\phi)\cos(ky\sin\phi)+F(\phi)\sin(kx\cos\phi)\sin(ky\sin\phi))d\phi
\label{A1}
\ee
where the functions $C(\phi),...,F(\phi)$ depend on an image and can be expanded into the following Fourier series
\be
X(\phi)=X_0+\sum_{p\geq 1}\ll(X_p^{(1)}\sin(p\phi)+X_p^{(2)}\cos(p\phi)\r)\nn\\
X=C,...,F
\label{A2}
\ee

Taking into account the following Fourier series for $e^{iz\cos\phi}$ \cite{9}
\be
e^{iz\cos\phi}=J_0(z)+2\sum_{r\geq 1}i^rJ_r(z)\cos(r\phi)
\label{A3}
\ee
where $J_r(z),\;r=0,1,2,...$ are the Bessel functions
we can expand also into the Fourier series the functions $\cos(kx\cos\phi),...,\;\sin(ky\sin\phi)$ present in the sum of \mref{A1} to get
\be
\cos(kx\cos\phi)=J_0(kx)+2\sum_{r\geq 1}(-1)^rJ_{2r}(kx)\cos(2r\phi)\nn\\
\sin(kx\cos\phi)=2\sum_{r\geq 1}(-1)^rJ_{2r-1}(kx)\cos((2r-1)\phi)\nn\\
\cos(ky\sin\phi)=J_0(ky)+2\sum_{r\geq 1}J_{2r}(ky)\cos(2r\phi)\nn\\
\sin(ky\sin\phi)=2\sum_{r\geq 1}J_{2r-1}(ky)\sin((2r-1)\phi)
\label{A4}
\ee

Using \mref{A2} and \mref{A3} in \mref{A1} we get therefore
\be
\Psi(x,y;k)=2\pi\ll(J_0(kx)J_0(ky)C_0+\sum_{r\geq 1}\ll(J_0(kx)J_{2r}(ky)+(-1)^rJ_0(ky)J_{2r}(kx)\r)C_{2r}^{(2)}+\r.\nn\\
                                    \sum_{r,s\geq 1}(-1)^rJ_{2s}(kx)J_{2r}(ky)\ll(C_{2r+2s}^{(2)}+C_{2|r-s|}^{(2)}\r)+\nn\\
                             J_0(kx)\sum_{r\geq 1}J_{2r-1}(ky)D_{2r-1}^{(1)}+\nn\\
                   \sum_{s\geq r\geq 1}(-1)^rJ_{2r}(kx)J_{2s+1}(ky)\ll(D_{2r+2s+1}^{(1)}+D_{2s-2r+1}^{(1)}\r)+\nn\\
                   \sum_{r>s\geq 0}(-1)^rJ_{2r}(kx)J_{2s+1}(ky)\ll(D_{2r+2s+1}^{(1)}-D_{2r-2s-1}^{(1)}\r)-\nn\\
                   J_0(ky)\sum_{r\geq 0}(-1)^rJ_{2r+1}(kx)E_{2r+1}^{(2)}-\nn\\
                   \sum_{r\geq 0,s\geq 1}(-1)^rJ_{2r+1}(kx)J_{2s}(ky)\ll(E_{2r+2s+1}^{(2)}+E_{|2r-2s-1|}^{(2)}\r)-\nn\\
                   \sum_{r>s\geq 0}(-1)^rJ_{2r+1}(kx)J_{2s+1}(ky)\ll(F_{2r+2s+2}^{(1)}-F_{2r-2s}^{(1)}\r)-\nn\\
                   \sum_{s>r\geq 0}(-1)^rJ_{2r+1}(kx)J_{2s+1}(ky)\ll(F_{2r+2s+2}^{(1)}+F_{2s-2r}^{(1)}\r)-\nn\\
                   \ll.\sum_{s\geq 0}(-1)^sJ_{2s+1}(kx)J_{2s+1}(ky)F_{4s+2}^{(1)}\r)
\label{A5}
\ee

Inspecting the coefficients $C_r^{(1)},...,F_r{(2)},\;r>0$ present in \mref{A5} one can notice that they can be collected into a single function $A(\phi)$
defined by \mref{6} by the following identification
\be
A_0=C_0,\;A_{2p}^{(1)}=F_{2p}^{(1)},\;\;A_{2p+1}^{(1)}=D_{2p+1}^{(1)},\;A_{2p}^{(2)}=C_{2p}^{(2)},\;A_{2p+1}^{(2)}=E_{2p+1}^{(2)}\nn\\
p=1,2,3,...
\label{A6}
\ee

A problem of convergence of the series like \mref{A5} is similar to the power one and can be expressed by the standard notion of the radius of
convergence by introducing the polar coordinates $x=r\cos\phi,\;y=r\sin\phi$ so that for the series $\sum_{m,n\geq 0}a_{m,n}x^my^n$ the respective
radius $\rho$ depends on $\phi$ and is given by
\be
\rho^{-1}(\phi)=\limsup_{n\to\infty}\ll|\sum_{m=0}^na_{m,n-m}\cos^m\phi\sin^{n-m}\phi\r|^{\frac{1}{n}}
\label{A7}
\ee
if one takes into account that $J_n(x)\sim x^n/2^nn!$ when $n\to\infty$.

Of course the boundary of the area $S$ of convergence of the power series $\sum_{m,n\geq 0}a_{m,n}x^my^n$ is then given by
$x=\rho(\phi)\cos\phi,\;y=\rho(\phi)\sin\phi$.

\section{Fourier series for SPS with discontinuities}

\hskip+2em The role of discontinuities of SPS on its RBRS has been partly taken into account on the several opportunities of calculations of the
Fourier series coefficients, see \mref{27a} and similar formulae in the paper. In fact discontinuities of SPS influence on the relations
between the coefficients of a given Fourier series and its derivatives. Let us illustrate these relations considering an example $\Psi(x)$,
of function of a
single variable $x$, $2\pi$-periodic, smooth on the segment $(0,2\pi)$ except the point $c,\;0<c<2\pi,$ where it is discontinuous having however left and
right derivatives. Assume its Fourier series to be
\be
\Psi^{FS}(x)=\sum_{j=1,2}\sum_{n\geq 0}X_{j,n}f_j(nx),\;\;\;i=1,2
\label{B1}
\ee
Then the coefficients $Y_{j,n}$ and $Z_{j,n}$ of the Fourier series of its respective first and second derivatives are
\be
Y_{j,n}=\int_0^{2\pi}\Psi'(x)f_j(nx)dx=(\Psi(c_+)-\Psi(c_-))f_j(nc)-n(-1)^{j+1}X_{j+1,n}\nn\\
Z_{j,n}=\int_0^{2\pi}\Psi''(x)f_j(nx)dx=(\Psi'(c_+)-\Psi'(c_-))f_j(nc)-n(-1)^{j+1}Y_{j+1,n}=\nn\\
              (\Psi'(c_+)-\Psi'(c_-))f_j(nc)+n(-1)^{j+1}(\Psi(c_+)-\Psi(c_-))f_{j+1}(nc)-n^2X_{j,n}
\label{B2}
\ee

If now $\Psi(x)$ is continuous at the point $x=c$ with the discontinuous first derivative then the respective Fourier series for its first and the second derivatives
are the following
\be
{\Psi'}^{FS}(x)=\sum_{j=1,2}(-1)^j\sum_{n\geq 0}nX_{j+1,n}f_j(nx)\nn\\
{\Psi''}^{FS}(x)=\sum_{j=1,2}\sum_{n\geq 0}((\Psi'(c_+)-\Psi'(c_-))f_j(nc)-n^2X_{j,n})f_j(nx)
\label{B3}
\ee
while in the opposite case we have
\be
{\Psi'}^{FS}(x)=\sum_{j=1,2}\sum_{n\geq 0}(\Psi(c_+)-\Psi(c_-))f_j(nc)-n(-1)^{j+1}X_{j+1,n})f_j(nx)\nn\\
{\Psi''}^{FS}(x)=\sum_{j=1,2}\sum_{n\geq 0}(n(-1)^{j+1}(\Psi(c_+)-\Psi(c_-))f_{j+1}(nc)-n^2X_{j,n})f_j(nx)
\label{B4}
\ee

\section{Definition of the quantities entering the formulae \mref{18a}}
\be
({\bf X}_{ij}^{(k)})_{mn}=X_{ij;mn}^{(k)},\;\;\;k=a,b,\;\;\;({\bf X}_{ij}^{P_3})_{mn}=X_{ij;mn}^{P_3}\nn\\
({\bf X}_{ij}^{(a,xy)})_{mn}=(-1)^{i+j}\frac{4\pi^2mn}{bc}({\bf X}_{ij}^{(a)})_{mn}\nn\\
({\bf X}_{ij}^{(a,x^2)})_{mn}=\frac{4\pi^2m^2}{b^2}({\bf X}_{ij}^{(a)})_{mn}\nn\\
({\bf X}_{ij}^{(a,y^2)})_{mn}=\frac{4\pi^2n^2}{c^2}({\bf X}_{ij}^{(a)})_{mn}\nn\\
({\bf X}_{ij}^{(b,xy)})_{mn}=(-1)^{i+j}\frac{4\pi^2mn}{ad}({\bf X}_{ij}^{(b)})_{mn}\nn\\
({\bf X}_{ij}^{(b,x^2)})_{mn}=\frac{4\pi^2m^2}{a^2}({\bf X}_{ij}^{(b)})_{mn}\nn\\
({\bf X}_{ij}^{(b,y^2)})_{mn}=\frac{4\pi^2n^2}{d^2}({\bf X}_{ij}^{(b)})_{mn}\nn\\
({\bf X}_{ij}^{P_3,xy})_{mn}=(-1)^{i+j}\frac{4\pi^2mn}{(a+b)c}({\bf X}_{ij}^{P_3})_{mn}\nn\\
({\bf X}_{ij}^{P_3,x^2})_{mn}=\frac{4\pi^2m^2}{(a+b)^2}({\bf X}_{ij}^{P_3})_{mn}\nn\\
({\bf X}_{ij}^{P_3,y^2})_{mn}=\frac{4\pi^2n^2}{c^2}({\bf X}_{ij}^{P_3})_{mn}\nn\\
(\tilde{\bf D}_{ij}^{(a,x)})_{mn}=
\frac{1}{bc}\oint_{L_1}dl\cos\gamma_a(l){\tilde\Psi}_a^{(ji)}(x'(l),y(l))f_i\ll(\pi m\frac{x'(l)}{b}\r)f_j\ll(\pi n\frac{y(l)}{c}\r)\nn\\
(\check{\tilde{\bf D}}_{ij}^{(a,x)})_{mn}=(-1)^i\frac{\pi m}{b}(\tilde{\bf D}_{ij}^{(a,x)})_{mn}\nn\\
(\tilde{\bf D}_{ij}^{(a,y)})_{mn}=
\frac{1}{bc}\oint_{L_1}dl\sin\gamma_a(l){\tilde\Psi}_a^{(ji)}(x'(l),y(l))f_i\ll(\pi m\frac{x'(l)}{b}\r)f_j\ll(\pi n\frac{y(l)}{c}\r)\nn\\
(\check{\tilde{\bf D}}_{ij}^{(a,y)})_{mn}=(-1)^j\frac{\pi n}{c}(\tilde{\bf D}_{ij}^{(a,y)})_{mn}\nn\\
(\breve{\tilde{\bf D}}_{ij}^{(a,y)})_{mn}=(-1)^i\frac{\pi m}{b}(\tilde{\bf D}_{ij}^{(a,y)})_{mn}\nn\\
(\tilde{\bf D}_{ij}^{(a,xy)})_{mn}=
\frac{1}{bc}\oint_{L_1}dl\cos\gamma_a(l)\frac{\p{\tilde\Psi}_a^{(ji)}(x'(l),y(l))}{\p y}f_i\ll(\pi m\frac{x'(l)}{b}\r)f_j\ll(\pi n\frac{y(l)}{c}\r)\nn\\
(\tilde{\bf D}_{ij}^{(a,x^2)})_{mn}=
\frac{1}{bc}\oint_{L_1}dl\cos\gamma_a(l)\frac{\p{\tilde\Psi}_a^{(ji)}(x'(l),y(l))}{\p x}f_i\ll(\pi m\frac{x'(l)}{b}\r)f_j\ll(\pi n\frac{y(l)}{c}\r)\nn\\
(\tilde{\bf D}_{ij}^{(a,y^2)})_{mn}=
\frac{1}{bc}\oint_{L_1}dl\sin\gamma_a(l)\frac{\p{\tilde\Psi}_a^{(ji)}(x'(l),y(l))}{\p y}f_i\ll(\pi m\frac{x'(l)}{b}\r)f_j\ll(\pi n\frac{y(l)}{c}\r)\nn\\
({\bf D}_{ij}^{(b,x)})_{mn}=
\frac{1}{ad}\oint_{L_2}dl\cos\gamma_b(l){\Psi}_b^{(ji)}(x(l),y(l))f_i\ll(\pi m\frac{x(l)}{a}\r)f_j\ll(\pi n\frac{y(l)}{d}\r)\nn\\
(\check{\bf D}_{ij}^{(b,x)})_{mn}=(-1)^i\frac{\pi m}{a}({\bf D}_{ij}^{(b,x)})_{mn}\nn\\
({\bf D}_{ij}^{(b,y)})_{mn}=
\frac{1}{ad}\oint_{L_2}dl\sin\gamma_b(l){\Psi}_b^{(ji)}(x(l),y(l))f_i\ll(\pi m\frac{x(l)}{a}\r)f_j\ll(\pi n\frac{y(l)}{d}\r)\nn\\
(\check{\bf D}_{ij}^{(b,y)})_{mn}=(-1)^j\frac{\pi n}{d}({\bf D}_{ij}^{(b,y)})_{mn}\nn\\
(\breve{\bf D}_{ij}^{(b,y)})_{mn}=(-1)^i\frac{\pi m}{a}({\bf D}_{ij}^{(b,y)})_{mn}\nn\\
({\bf D}_{ij}^{(b,xy)})_{mn}=
\frac{1}{ad}\oint_{L_2}dl\cos\gamma_b(l)\frac{\p\Psi_b^{(ji)}(x(l),y(l))}{\p y}f_i\ll(\pi m\frac{x(l)}{a}\r)f_j\ll(\pi n\frac{y(l)}{d}\r)\nn\\
({\bf D}_{ij}^{(b,x^2)})_{mn}=
\frac{1}{ad}\oint_{L_2}dl\cos\gamma_b(l)\frac{\p\Psi_b^{(ji)}(x(l),y(l))}{\p x}f_i\ll(\pi m\frac{x(l)}{a}\r)f_j\ll(\pi n\frac{y(l)}{d}\r)\nn\\
({\bf D}_{ij}^{(b,y^2)})_{mn}=
\frac{1}{ad}\oint_{L_2}dl\sin\gamma_b(l)\frac{\p\Psi_b^{(ji)}(x(l),y(l))}{\p y}f_i\ll(\pi m\frac{x(l)}{a}\r)f_j\ll(\pi n\frac{y(l)}{d}\r)\nn\\
({\bf D}_{ij}^{P_3,x})_{mn}=\nn\\
\frac{1}{(a+b)c}\oint_{L_3}dl\cos\gamma_{P_3}(l)\Psi_{P_3}^{(ji)}(x(l),y(l))f_i\ll(\pi m\frac{x(l)}{a+b}\r)f_j\ll(\pi n\frac{y(l)}{c}\r)\nn\\
(\check{\bf D}_{ij}^{P_3,x})_{mn}=(-1)^i\frac{\pi m}{a+b}({\bf D}_{ij}^{P_3,x})_{mn}\nn\\
({\bf D}_{ij}^{P_3,y})_{mn}=\nn\\
\frac{1}{(a+b)c}\oint_{L_3}dl\sin\gamma_{P_3}(l)\Psi_{P_3}^{(ji)}(x(l),y(l))f_i\ll(\pi m\frac{x(l)}{a+b}\r)f_j\ll(\pi n\frac{y(l)}{c}\r)\nn\\
(\check{\bf D}_{ij}^{P_3,y})_{mn}=(-1)^j\frac{\pi n}{c}({\bf D}_{ij}^{P_3,y})_{mn}\nn\\
(\breve{\bf D}_{ij}^{P_3,y})_{mn}=(-1)^i\frac{\pi m}{a+b}({\bf D}_{ij}^{P_3,y})_{mn}\nn\\
({\bf D}_{ij}^{P_3,xy})_{mn}=\nn\\
\frac{1}{(a+b)c}\oint_{L_3}dl\cos\gamma_{P_3}(l)\frac{\p\Psi_{P_3}^{(ji)}(x(l),y(l))}{\p y}f_i\ll(\pi m\frac{x(l)}{a+b}\r)f_j\ll(\pi n\frac{y(l)}{c}\r)\nn\\
({\bf D}_{ij}^{P_3,x^2})_{mn}=\nn\\
\frac{1}{(a+b)c}\oint_{L_3}dl\cos\gamma_{P_3}(l)\frac{\p\Psi_{P_3}^{(ji)}(x(l),y(l))}{\p x}f_i\ll(\pi m\frac{x(l)}{a+b}\r)f_j\ll(\pi n\frac{y(l)}{c}\r)\nn\\
({\bf D}_{ij}^{P_3,y^2})_{mn}=\nn\\
\frac{1}{(a+b)c}\oint_{L_3}dl\sin\gamma_{P_3}(l)\frac{\p\Psi_{P_3}^{(ji)}(x(l),y(l))}{\p y}f_i\ll(\pi m\frac{x(l)}{a+b}\r)f_j\ll(\pi n\frac{y(l)}{c}\r)\nn\\
(\alpha_{ij})_{mn}=\frac{1}{a+b}\int_{-a-2b}^{-a}dxf_i\ll(\pi m\frac{x}{a+b}\r)f_j\ll(\pi n\frac{x}{b}\r)\nn\\
(\beta_{ij})_{mn}=\frac{1}{a+b}\int_{-a}^adxf_i\ll(\pi m\frac{x}{a+b}\r)f_j\ll(\pi n\frac{x}{b}\r)\nn\\
(\gamma_{ij})_{mn}=\frac{1}{c}\int_{-c}^cdyf_i\ll(\pi m\frac{y}{d}\r)f_j\ll(\pi n\frac{y}{c}\r)
\label{C1}
\ee

\section{Stationary energy spectra in RB as homogeneous functions of periods}

\hskip+2em The basic general property which can be inferred from the presented in the previous sections results is that the discrete energy spectra in RB
are homogeneous functions of periods which they depend on. To see this let us rescale the two dimensions of RBRS $r$-times. It means that SPS
$\Psi_n(x,y;D_1,...,D_{2g};k_n(D_1,...,D_{2g}))$ is also rescaled to \[\Psi_n(x,y;rD_1,...,rD_{2g};k_n(rD_1,...,rD_{2g}))\] the latter satisfying the
following SE
\be
\ll(\frac{\p^2}{\p x^2}+\frac{\p^2}{\p y^2}\r)\Psi_n(x,y;rD_1,...,rD_{2g};k_n(rD_1,...,rD_{2g}))+\nn\\
k_n^2(rD_1,...,rD_{2g})\Psi_n(x,y;rD_1,...,rD_{2g};k_n(rD_1,...,rD_{2g}))=0
\label{D1}
\ee

But as a result of the rescaling we should have also
\be
\Psi_n(rx,ry;rD_1,...,rD_{2g};k_n(rD_1,...,rD_{2g}))=e^{i\alpha}\Psi_n(x,y;D_1,...,D_{2g};k_n(D_1,...,D_{2g}))
\label{D2}
\ee

Calculating the Laplasian of both the sides in \mref{D2} we get
\be
r^2k_n^2(rD_1,...,rD_{2g})=k_n^2(D_1,...,D_{2g})
\label{D3}
\ee
i.e.
\be
k_n(rD_1,...,rD_{2g})=r^{-1}k_n(D_1,...,D_{2g})
\label{D4}
\ee

\end{document}